\documentclass[12pt]{article}
\usepackage{epsfig}
\usepackage{amssymb}
\usepackage{cite}
\usepackage{authblk}

\begin{document}

\begin{titlepage}

\title{\bf Strange quark contributions to nucleon mass and spin
from lattice QCD}
\author{M.~Engelhardt}

\affil{\em Department of Physics, New Mexico State University,\\
\em Las Cruces, NM 88003, USA}

\maketitle

\begin{abstract}
Contributions of strange quarks to the mass and spin of the nucleon,
characterized by the observables $f_{T_s } $ and $\Delta s$,
respectively, are investigated within lattice QCD. The calculation
employs a 2+1-flavor mixed-action lattice scheme, thus treating the
strange quark degrees of freedom in dynamical fashion.
Numerical results are obtained at three pion masses,
$m_{\pi } = 495\, \mbox{MeV} $, $356\, \mbox{MeV} $, and $293\, \mbox{MeV} $,
renormalized, and chirally extrapolated to the physical pion mass.
The value extracted for $\Delta s$ at the physical pion mass in
the $\overline{MS} $ scheme at a scale of
$2\, \mbox{GeV} $ is $\Delta s = -0.031(17)$, whereas the
strange quark contribution to the nucleon mass amounts to
$f_{T_s } =0.046(11)$. In the employed mixed-action scheme, the
nucleon valence quarks as well as the strange quarks entering
the nucleon matrix elements which determine $f_{T_s } $ and $\Delta s$
are realized as domain wall fermions, propagators of which are evaluated
in MILC 2+1-flavor dynamical asqtad quark ensembles. The use of domain
wall fermions leads to mild renormalization behavior which proves
especially advantageous in the extraction of $f_{T_s } $.
\end{abstract}

PACS: 12.38.Gc, 14.20.Dh

\end{titlepage}

\setcounter{page}{2}

\section{Introduction}
Strange quarks represent the lightest quark flavor not present in the
valence component of the nucleon. Their study can thus provide insight
into sea quark effects in the nucleon in isolated fashion. The two most
fundamental properties of the nucleon are its mass and spin. The
investigation presented here focuses on the extent to which those two
properties are influenced by the strange quark degrees of freedom.
The strange contributions to nucleon mass and spin can be characterized by
the matrix elements
\begin{equation}
f_{T_s } =\frac{m_s }{m_N } \left[
\langle N | \int d^3 y\, \bar{s} s | N\rangle
-\langle 0 | \int d^3 y\, \bar{s} s | 0\rangle \right]
\label{matel1}
\end{equation}
and
\begin{equation}
\Delta s = \langle N,j | \int d^3 y\, \bar{s} \gamma_{j} \gamma_{5} s
| N,j \rangle
\label{matel2}
\end{equation}
respectively, where $| N,j \rangle $ denotes a nucleon state with spin
polarized in the $j$-direction. In the case of the scalar matrix element,
the vacuum expectation value, i.e., the vacuum strange scalar condensate, is
subtracted; the intention is, of course, to measure the strangeness content
of the nucleon {\em relative} to the vacuum. In the case of the axial
matrix element, no subtraction is necessary since the corresponding vacuum
expectation value vanishes. Note that $\Delta s$ measures specifically
the contribution of strange quark spin to nucleon spin; strange quark
angular momentum constitutes a separate contribution not considered here.

Aside from representing a fundamental characteristic of the nucleon
in its own right, the scalar strange content $f_{T_s } $ is also an
important parameter in the context of dark matter searches
\cite{bottino,ellis1,ellis2,bertone}. Assuming that the coupling
of dark matter to baryonic matter is mediated by the Higgs field,
the dark matter detection rate depends sensitively on the quark scalar
matrix elements in the nucleon, cf., e.g, the neutralino-nucleon scalar
cross-section considered in \cite{bottino}. One a priori reasonable
scenario is that the strange quark furnishes a particularly favorable
channel \cite{bottino}, since, on the one hand, it features a much larger
Yukawa coupling to the Higgs field than the light quarks, and, on the
other hand, is not so heavy as to be only negligibly represented in the
nucleon's sea quark content. As a consequence, an accurate estimate of
$f_{T_s } $ is instrumental in assessing the discovery potential for
dark matter candidates.

The contribution of strange quark spin to nucleon spin $\Delta s$
is, in principle, more directly accessible to experiment than $f_{T_s } $.
$\Delta s$ represents the first moment of the strange quark helicity
distribution $\Delta s(x)$ (including both quarks and antiquarks) as a
function of the momentum fraction $x$. The helicity distribution can be
determined via inclusive deep inelastic scattering and semi-inclusive deep
inelastic scattering \cite{herm1,herm2,compass1}. However, its extraction
in practice still has to rely to a certain extent on assumptions about the
dependence of $\Delta s(x)$ on $x$, even in the semi-inclusive channels
(which furnish direct information on $\Delta s(x)$), because of the
limitations in accessing small $x$ experimentally. Complementary information
about $\Delta s$ is obtained from the strange axial form factor of the
nucleon $G^{s}_{A} (Q^2 )$, which can be extracted by combining data from
parity-violating elastic electron-proton scattering and elastic
neutrino-proton scattering \cite{pate}. Extrapolation to zero momentum
transfer, $Q^2 =0$, again yields an estimate of $\Delta s$. Depending on
the specific extrapolations and/or model assumptions adopted in
determining $\Delta s$ via the various aforementioned avenues, both
significantly negative values for $\Delta s$ have been put forward
\cite{herm1,compass2}, as well as values compatible with zero \cite{herm2}.
An independent determination of $\Delta s$ via lattice QCD, as undertaken
in the present work, thus can be useful in several ways. Apart from
shedding light on the fundamental question of the decomposition of
nucleon spin, it can contribute constraints to phenomenological fits
of polarized parton distribution functions. Furthermore, it influences
spin-dependent dark matter cross sections \cite{ellis2}; although more
accurate determinations of the scalar matrix elements discussed further
above constitute the most urgent issue in reducing hadronic uncertainties
in dark matter searches, $\Delta s$ also plays a significant role in that
context.

A number of lattice QCD investigations of strange quark degrees of freedom
in the nucleon have recently been undertaken\cite{bali1,bali2,bali3,
jlqcd1,jlqcd2,jlqcd3,freeman1,freeman2,kfliu1,kfliu2,kfliu3,duerr,
drach,qcdsf,babich,young1,young2}, the majority of which have
focused specifically on the scalar content. Studies of the latter have
proceeded via two avenues: On the one hand, one can directly determine
the matrix element $\langle N|\bar{s} s|N\rangle $ via the appropriate
disconnected three-point function; this methodology was adopted
in\footnote{Strictly speaking, the method adopted in \cite{freeman1,freeman2}
constitutes a hybrid of the two methods.}
\cite{bali1,bali3,jlqcd2,jlqcd3,freeman1,freeman2,kfliu2,drach,babich}
and also in the present work, as described in detail further
below. A study of techniques suited to improve the efficiency of
this approach has been presented in \cite{cyprus}. On the other hand,
a somewhat less direct inference of the scalar strange quark content of
the nucleon is possible via the study of the baryon spectrum, which is
related via the Feynman-Hellmann theorem
\begin{equation}
\langle B | \bar{q} q | B\rangle = \frac{\partial m_B }{\partial m_q }
\end{equation}
to the corresponding sigma terms for the baryon state $|B\rangle $ and
quark flavor $q$. This avenue has been pursued in 
\cite{jlqcd1,jlqcd3,duerr,qcdsf}, and a related methodology, combining
lattice hadron spectrum data with chiral perturbation theory, was pursued
in \cite{young1,young2}.

The characteristics of these various investigations of the scalar strange
quark content of the nucleon are diverse. They include $N_f =2$
calculations, in which the strange quark degrees of freedom are
quenched \cite{bali3,jlqcd1,jlqcd2,babich}, but also $N_f =2+1$
\cite{bali1,jlqcd3,freeman1,freeman2,kfliu2,duerr,qcdsf,young1,young2}
and even $N_f =2+1+1$ \cite{freeman2,drach} calculations. In some
cases, lattice data at only one pion mass have been obtained to date
and no extrapolation to the physical point has been attempted. The most
stringent results obtained at the physical point including fully
dynamical strange quarks were reported in \cite{jlqcd3,freeman2,young2}.
Ref.~\cite{freeman2} quotes $m_N f_{T_s } /m_s = 0.637(55)(74)$ in the
$N_f =2+1$ case, and $m_N f_{T_s } /m_s =0.44(8)(5)$ in the $N_f =2+1+1$
case; translated to $f_{T_s } $ itself using $m_s =95\, \mbox{MeV} $
in the $\overline{MS} $ scheme at a scale of $2\, \mbox{GeV} $
\cite{pdg}, these correspond to $f_{T_s } = 0.064(6)(7)$ for $N_f =2+1$
and $f_{T_s } = 0.044(8)(5)$ for $N_f =2+1+1$. On the other hand,
\cite{jlqcd3} and \cite{young2} report significantly lower values.
In \cite{jlqcd3}, both direct three-point function calculations as
well as indirect determinations via the Feynman-Hellmann theorem are
used to arrive at a bound $f_{T_s } =0.009(15)(16)$. In \cite{young2},
a more indirect analysis using lattice hadron spectrum data and chiral
perturbation theory yields $m_N f_{T_s } = 21(6) \, \mbox{MeV} $, which
translates to $f_{T_s } =0.022(6)$. The results obtained in the
present work are of comparable accuracy, $f_{T_s } = 0.046(11)$,
cf.~(\ref{ftsfinal}), and are more consistent with the higher values
reported in \cite{freeman2}.

The strange axial matrix element $\Delta s$, on the other hand, has also
been investigated in \cite{bali1,bali2,babich}. Apart from the exploratory
study \cite{bali1}, which, however, did not attempt to renormalize the
results nor extrapolate them to the physical pion mass, these investigations
were based on dynamical quark ensembles containing only the two light
flavors in the quark sea; the present lattice
investigation, on the other hand, employs $N_f =2+1$ gauge ensembles,
thus treating the strange quark degrees of freedom in dynamical fashion.
The numerical results for $\Delta s$ obtained on this basis are
renormalized and chirally extrapolated, yielding the estimate
$\Delta s = -0.031(17)$, cf.~(\ref{dsfinal}), at the physical point
in the $\overline{MS} $ scheme at a scale of $2\, \mbox{GeV} $.
Within the uncertainties, this nevertheless remains compatible with
the values obtained in the aforementioned other studies, though it
is about 50\% larger in magnitude. This suggests that systematic
adjustments to the results quoted in those works through
unquenching of the strange quark degrees of freedom, renormalization,
and chiral extrapolation are not severe.

Aside from the two quantities $f_{T_s } $ and $\Delta s$ considered in
the present work, lattice investigations of the strange quark structure
within the nucleon have also considered generalizations to non-zero
momentum transfer, i.e., form factors \cite{kfliu1,babich}, including
calculations of the strange electric and magnetic form factors,
which are of interest in the context of corresponding experimental
efforts employing parity-violating electron-proton scattering \cite{emff}.
Furthermore, also the strange quark momentum fraction and strange quark
angular momenta in the nucleon have been investigated \cite{kfliu3}.

The present lattice investigation, a preliminary account of which was
given in \cite{latt10}, is based on a mixed-action scheme developed and
employed extensively by the LHP Collaboration \cite{lhpc07,lhpc08,lhpc10}.
The nucleon valence quarks as well as the strange quark fields appearing in
the operator insertions in eqs.~(\ref{matel1}) and (\ref{matel2}) are
realized as domain wall fermions, propagators of which are evaluated in
the background of (HYP-smeared) 2+1-flavor dynamical asqtad quark ensembles
provided by the MILC Collaboration. Though computationally expensive,
domain wall fermions lead to benign renormalization properties which
prove especially advantageous in the case of the $f_{T_s } $ observable;
the substantial systematic uncertainties due to strong additive quark
mass renormalizations encountered in analogous calculations using
Wilson fermions \cite{bali3,babich} are avoided. The computational
scheme is described in detail in section~\ref{schemesec}.
Section~\ref{resultsec} provides the raw numerical results, the
renormalization of which is discussed in section~\ref{renormsec}.
The renormalized results are extrapolated to the physical pion mass
in section~\ref{chisec}, and systematic uncertainties and adjustments
are considered in section~\ref{syssec}, with conclusions presented in
section~\ref{concsec}.

\section{Computational scheme}
\label{schemesec}
\subsection{Correlator ratios}
\begin{figure}[h]
\vspace{0.5cm}
\begin{center}
\includegraphics[width=6cm]{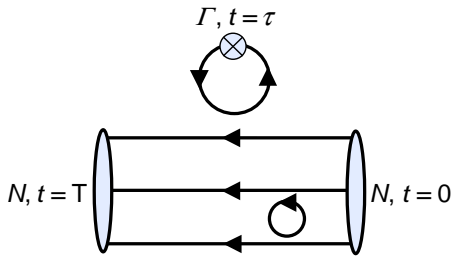}
\end{center}
\vspace{-1.2cm}
\caption{Disconnected contribution to nucleon matrix elements. The nucleon
propagates between a source at $t=0$ and a sink at $t=T$; the
insertion of $\Gamma \equiv \Gamma^{obs} $ occurs at an intermediate time
$t=\tau $.}
\label{ddiagram}
\end{figure}

The lattice calculation of the nucleon matrix elements in (\ref{matel1}),
(\ref{matel2}) proceeds in standard fashion via correlator ratios of the
type
\begin{equation}
R[\ \Gamma^{nuc} , \Gamma^{obs} \ ] (\tau ,T)
= \frac{\left\langle \ \left[
\Gamma^{nuc}_{\alpha \beta } \ \Sigma_{\vec{x} } \
N_{\beta } (\vec{x} ,T) \bar{N}_{\alpha } (0,0) \right]
\cdot \left[ - \Gamma^{obs}_{\gamma \rho }
\Sigma_{\vec{y} } \ s_{\rho } (\vec{y} ,\tau )
\bar{s}_{\gamma } (\vec{y} ,\tau ) \right] \
\right\rangle }{\left\langle \
\Gamma^{unpol}_{\alpha \beta } \ \Sigma_{\vec{x} } \
N_{\beta } (\vec{x} ,T) \bar{N}_{\alpha } (0,0) \ \right\rangle }
\label{corrrat}
\end{equation}
with nucleon interpolating fields $N$ of the standard form
(quoting here, for definiteness in flavor structure, the proton case)
\begin{equation}
N_{\gamma } = \epsilon_{abc} u^c_{\gamma } u^a_{\alpha }
\left( C\gamma_{5} \right)_{\alpha \beta } d^b_{\beta }
\label{nucinp}
\end{equation}
where $C$ denotes the charge conjugation matrix.
The sums over spatial position $\vec{x} $ project the nucleon states onto
zero momentum, whereas the sum over spatial position $\vec{y} $ is simply
transcribed from (\ref{matel1}), (\ref{matel2}). Since the nucleon contains
no strange valence quarks, the three-point function averaged in the
numerator of (\ref{corrrat}) factorizes, as written, into the nucleon
two-point function and the strange quark loop. I.e., only the disconnected
diagram, cf.~Fig.~\ref{ddiagram}, contributes to the matrix elements under
consideration.
The strange quark fields have already been reordered in the numerator
of (\ref{corrrat}) such as to make the standard minus sign associated
with quark loops explicit. Finally, the $\Gamma $ matrices allow one to
choose the appropriate nucleon polarization and strange quark operator
insertion structures. The denominator of (\ref{corrrat}) corresponds to
the unpolarized nucleon two-point function, obtained using
\begin{equation}
\Gamma^{unpol} = \frac{1+\gamma_{4} }{2} \ .
\end{equation}
In the numerator, for the purpose of evaluating $f_{T_s } $, unpolarized
nucleon states are appropriate, corresponding to the choice
$\Gamma^{nuc} = \Gamma^{unpol} $; furthermore, the scalar strange quark
insertion is obtained by choosing $\Gamma^{obs} = 1$,
\begin{equation}
\frac{m_s }{m_N } \left( R[\ \Gamma^{unpol} , 1\ ] (\tau ,T)
\ - \ [\mbox{VEV}] \right)
\ \equiv \ R\{ f_{T_s } \} \
\stackrel{T \gg \tau \gg 0}{\longrightarrow} \ f_{T_s }
\label{corrfts}
\end{equation}
with the vacuum expectation value
\begin{equation}
[\mbox{VEV}] = \langle -\Sigma_{\vec{y} } \
s_{\gamma } (\vec{y} ,\tau ) \bar{s}_{\gamma } (\vec{y} ,\tau ) \rangle
\end{equation}
to be subtracted.

On the other hand, $\Delta s$ is obtained by using the projector onto
nucleon states polarized in the positive/negative $j$-direction in the
numerator of (\ref{corrrat}),
\begin{equation}
\Gamma^{nuc} = \frac{1\mp i\gamma_{j} \gamma_{5} }{2} \Gamma^{unpol} \ ,
\end{equation}
as well as the operator insertion structure
$\Gamma^{obs} = \gamma_{j} \gamma_{5} $. Averaging over positive/negative
$j$-direction (with a relative minus sign) as well as the three spatial
$j$ to improve statistics leads one to evaluate
\begin{equation}
-i\, \cdot 2 \cdot \frac{1}{3} \sum_{j=1}^{3}
R[\ (-i\gamma_{j} \gamma_{5} /2) \ \Gamma^{unpol} ,
\gamma_{j} \gamma_{5} \ ] (\tau ,T)
\ \equiv \ R\{ \Delta s \} \
\stackrel{T \gg \tau \gg 0}{\longrightarrow} \ \Delta s \ ,
\label{corrds}
\end{equation}
where the prefactor $2$ compensates for the fact that the ratio
(\ref{corrrat}) is normalized using the unpolarized nucleon two-point
function in the denominator, even when the numerator is restricted to a
particular polarization. Lastly, the prefactor $(-i)$ cancels the
additional factor $i$ which the $\gamma_{5} $-matrix in the operator
insertion acquires when the calculation is cast in terms of Euclidean
lattice correlators; thus, the $\Delta s $ obtained through (\ref{corrds})
is already Wick-rotated back to Minkowski space-time. Note that
(\ref{corrds}) does not call for the subtraction of a vacuum expectation
value, since the latter vanishes in the ensemble average. In practice,
this numerical zero was nevertheless subtracted from (\ref{corrds}) in
order to reduce statistical fluctuations.

\subsection{Lattice setup}
\label{setupsec}
The averages in the correlator ratios (\ref{corrrat}) were carried out using
the three MILC $2+1$-flavor dynamical asqtad quark ensembles listed in
Table~\ref{tabmilc}. HYP-smearing was applied to the configurations.
\begin{table}[h]
\begin{center}
\begin{tabular}{|c|c|c|c|c|c|c|}
\hline
$am_{l}^{asq} $ & $am_{s}^{asq} $ & $am_{l}^{DWF} $ & $am_{s}^{DWF} $ &
 $ m_{\pi } $ & \# configs & $m_N $ \\
\hline \hline
0.007 & 0.05 & 0.0081 & 0.081 & 293 MeV & 468 & 1107 MeV \\
\hline
0.01 & 0.05 & 0.0138 & 0.081 & 356 MeV & 448 & 1155 MeV \\
\hline
0.02 & 0.05 & 0.0313 & 0.081 & 495 MeV & 486 & 1288 MeV \\
\hline
\end{tabular}
\end{center}
\caption{$N_f =2+1$, $20^3 \times 64 $ MILC asqtad ensembles with lattice
spacing $a=0.124\, \mbox{fm} $ used in the present investigation.
Uncertainties in the pion and nucleon masses extracted from the
corresponding two-point functions are under $1\% $.}
\label{tabmilc}
\end{table}
Both the valence quarks in the nucleon two-point functions and the
strange quark fields appearing in the matrix elements (\ref{matel1})
and (\ref{matel2}) were implemented using the domain wall fermion
discretization, with parameters $L_5 =16$, $M_5 =1.7$. The domain
wall quark masses, also listed in Table~\ref{tabmilc}, are fixed by
the requirement of reproducing the pion masses corresponding to the
MILC ensembles \cite{lhpc07}. This mixed action setup has been
employed extensively for studies of hadron structure by LHPC
\cite{lhpc07,lhpc10} and further details concerning its tuning
can be found in the mentioned references.

\begin{figure}
\begin{center}
\includegraphics[angle=-90,width=11.1cm]{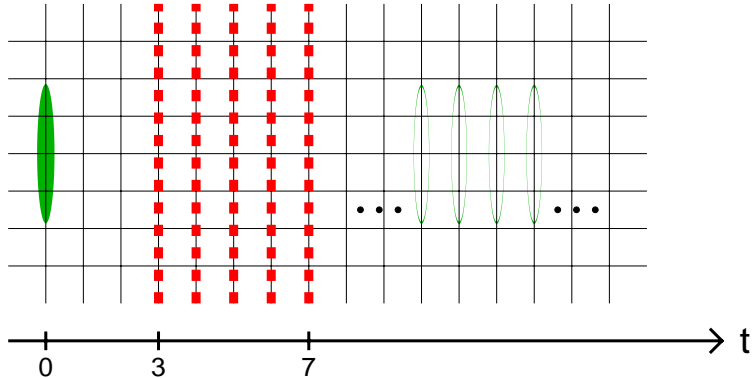}
\end{center}
\vspace{-1.1cm}
\caption{Setup of the lattice calculation. The nucleon source is located
at lattice time $t=0$. An average over operator insertion times $\tau $
is performed in the range $\tau =3,\ldots ,7$; accordingly, stochastic
sources are distributed over the bulk of the lattice in this entire time
range. The temporal position $T$ of the nucleon sink is variable.}
\label{setup}
\end{figure}

The space-time layout of the calculation is shown in Fig.~\ref{setup}.
The strange quark loop trace was evaluated using stochastic estimation.
Positioning the nucleon source at time $t=0$, bulk complex $Z(2)$
stochastic sources with support in all of space within the temporal range
$t=3,\ldots ,7$ were introduced. This corresponds to averaging the
correlator ratio (\ref{corrrat}) with respect to the operator insertion
time $\tau $ over the aforementioned time slices (after having duly divided
out the length of the temporal range). The stochastic estimate of the strange
quark loop trace was performed employing 1200 of the described stochastic
source vectors per gauge sample. In particular, obtaining a signal for
$\Delta s$ depends on accumulating high statistics in the stochastic
estimator. The scalar matrix element, on the other hand, requires less
statistics in the strange quark loop trace, but is more susceptible
to gauge fluctuations. The sink time $T$ at which the nucleon two-point
functions are contracted and projected onto zero momentum remains variable
in this layout.

While this scheme provides a high amount of averaging in relation to the
number of strange quark propagator inversions, it precludes testing for
a plateau in the three-point function by varying the operator insertion
time $\tau $. The positioning of the bulk stochastic source was motivated
by previous work \cite{lhpc07,lhpc10} in the same mixed-action
scheme, which investigated (connected contributions to) a wide variety
of observables, and the results of which indicate that, having evolved
in time from $t=0$ to $t=3$, excited state contaminations in the system
are already small compared to the statistical uncertainty of the
calculation performed here. To achieve this suppression of
excited state contributions, the quark fields in the nucleon sources
(\ref{nucinp}) are Wuppertal-smeared such as to optimize the overlap with
the nucleon ground state \cite{lhpc07,lhpc08,lhpc10}. Despite the
relative positioning of the nucleon source and the strange quark insertion
time range being fixed in this manner, it does nevertheless remain
possible to exhibit the extent to which results depend on the separation
between operator insertion and nucleon sink time $T$, which is still
variable in the present scheme. Correlator ratios will be
shown further below as a function of $T$, with asymptotic
behavior being seen to emerge three lattice time steps beyond the end
of the operator insertion range, i.e., for sink times $T\ge 10$. This
corroborates the suitability of the choice for the bulk stochastic source
time range adopted in the present calculation.

The correlator ratio (\ref{corrrat}) exhibits statistical fluctuations not
only due to the strange quark loop factor discussed above, but also due to
the nucleon two-point function factors. To reduce these fluctuations, it is
equally necessary to sample the latter to a sufficient extent. To this
end, multiple samples of the nucleon two-point function were obtained
by employing eight different spatial positions for the nucleon source on
the source time slice. In addition to this eight-fold sampling of the
nucleon two-point function, the described scheme can be accommodated
several times in temporally well-separated regions on the lattice;
in practice, three replicas of the entire setup specified above were
placed on the lattice, separated by 16 lattice spacings in the time
direction, thus further enhancing statistics. Each gauge configuration
therefore yielded altogether 24 samples of the numerator and denominator
in the correlator ratio (\ref{corrrat}). Note again that averaging was
further improved in the case of $\Delta s$ by taking into account nucleon
polarization axes aligned with all three coordinate axes, as already made
explicit in (\ref{corrds}).

\section{Numerical results}
\label{resultsec}
\begin{figure}
\centerline{\includegraphics[angle=-90,width=9cm]{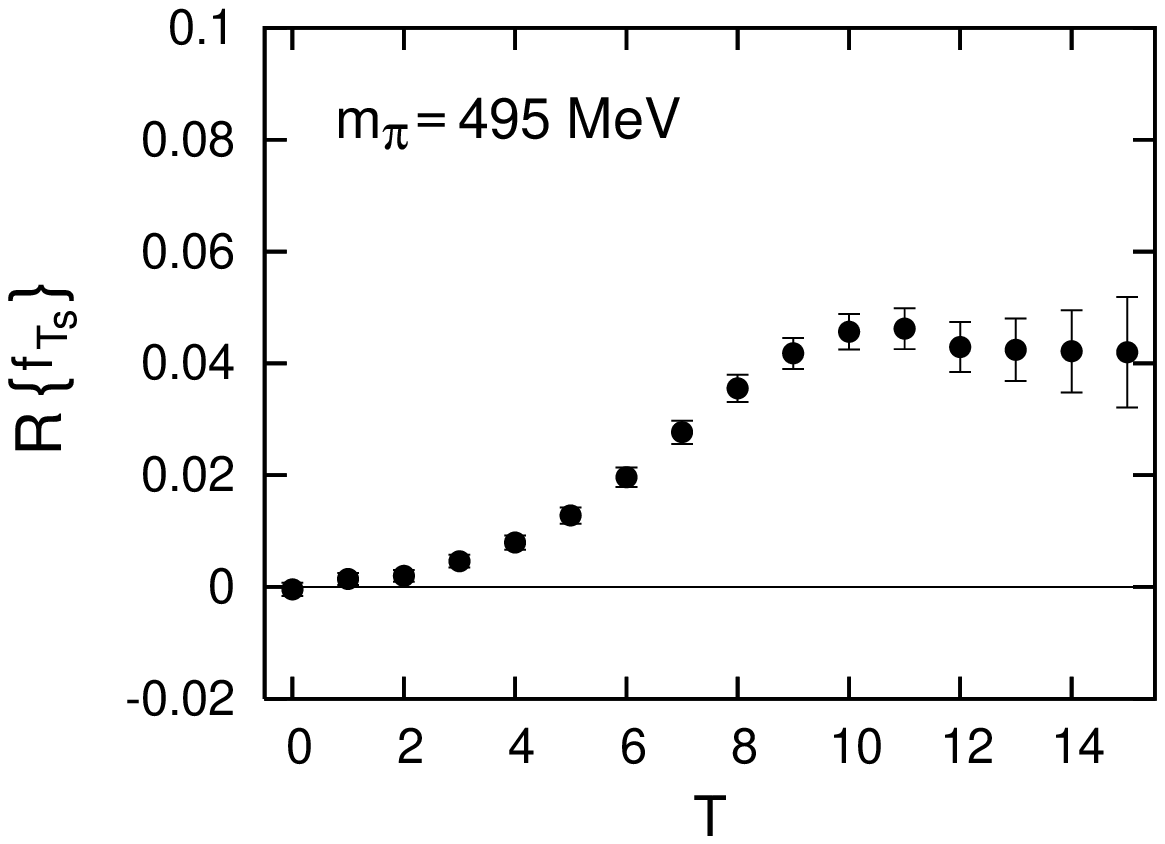} }
\vspace{0.3cm}
\centerline{\includegraphics[angle=-90,width=9cm]{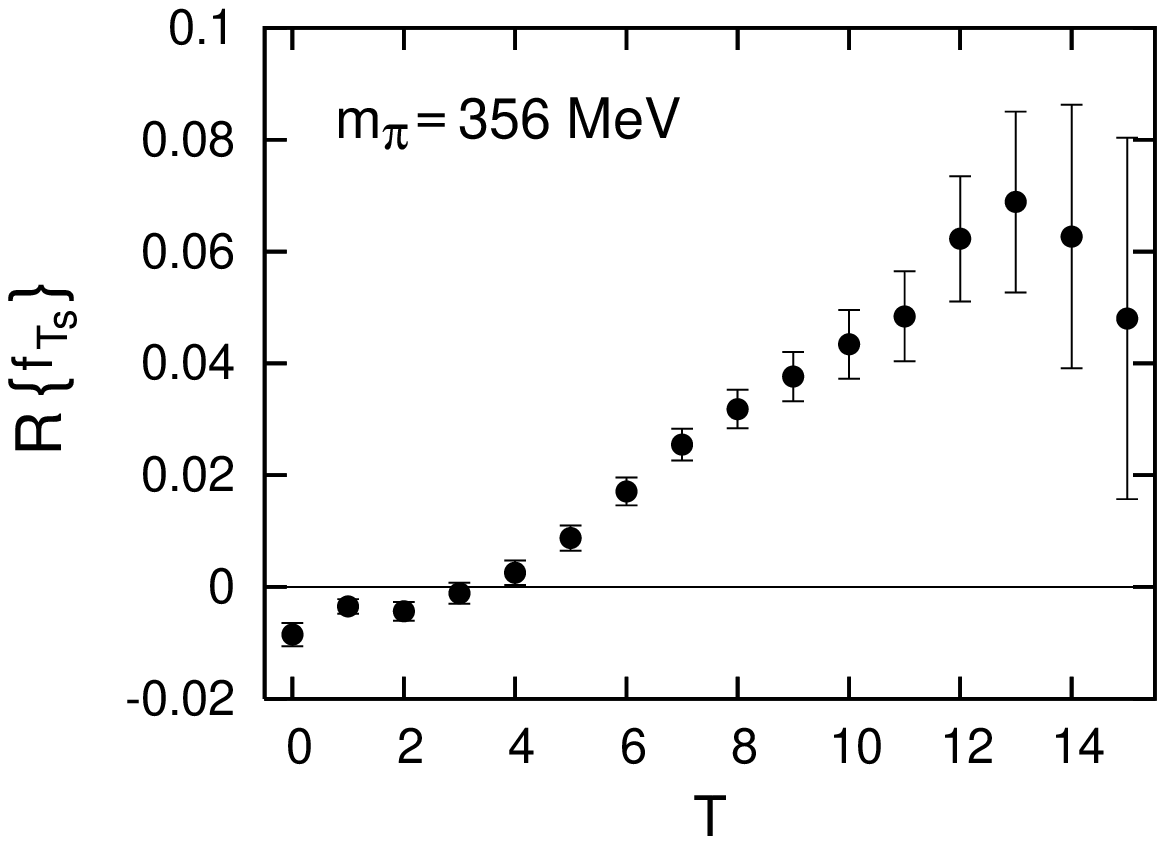} }
\vspace{0.3cm}
\centerline{\includegraphics[angle=-90,width=9cm]{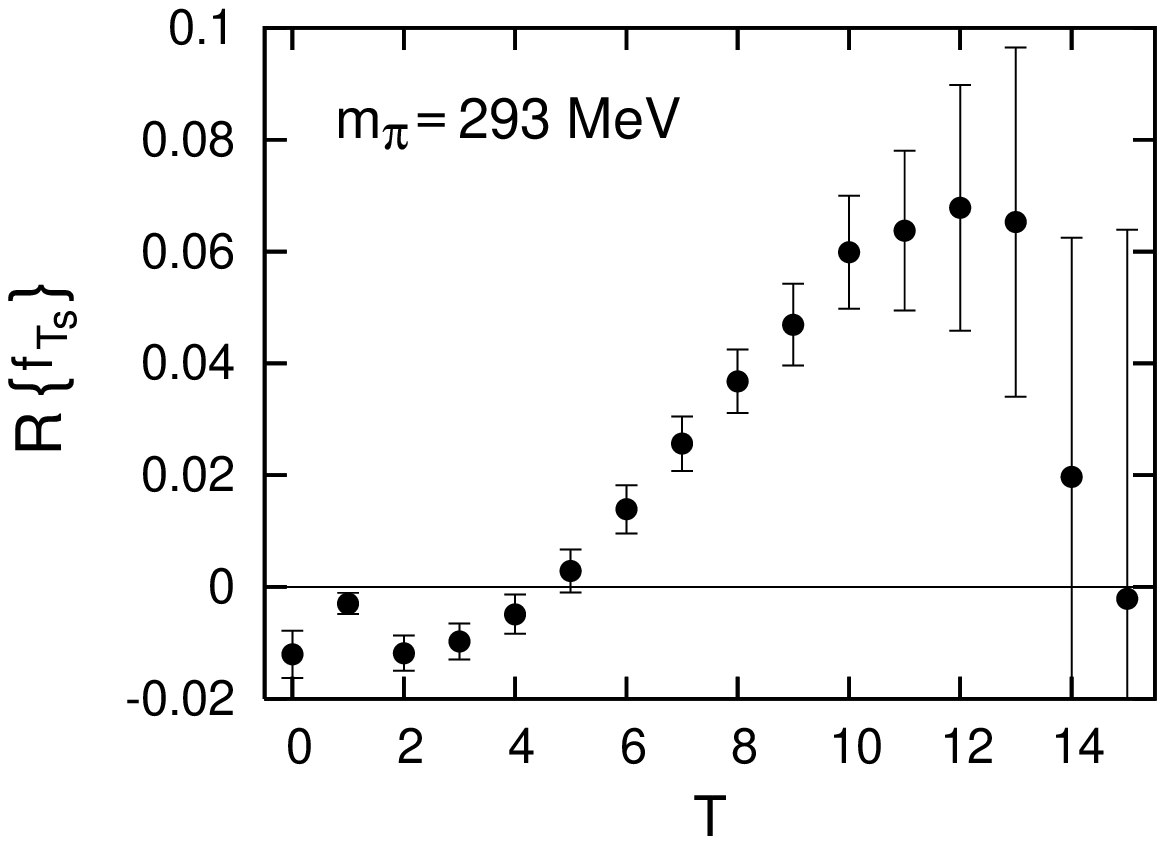} }
\vspace{0.3cm}
\caption{Correlator ratio $R\{ f_{T_s } \} $, cf.~(\ref{corrfts}),
averaged over insertion time $\tau $ as described in section \ref{setupsec},
as a function of sink time $T$, for the three pion masses considered.}
\label{resfts}
\end{figure}

\begin{figure}
\centerline{\includegraphics[angle=-90,width=9cm]{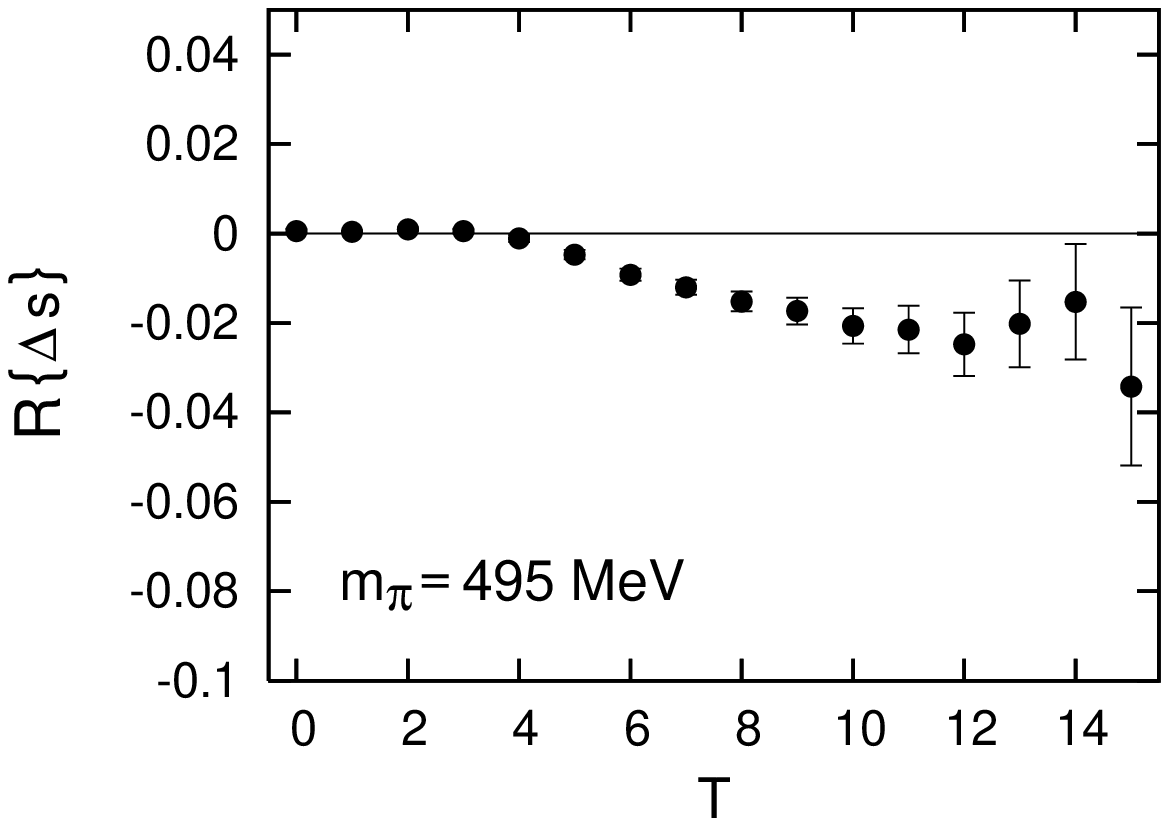} }
\vspace{0.3cm}
\centerline{\includegraphics[angle=-90,width=9cm]{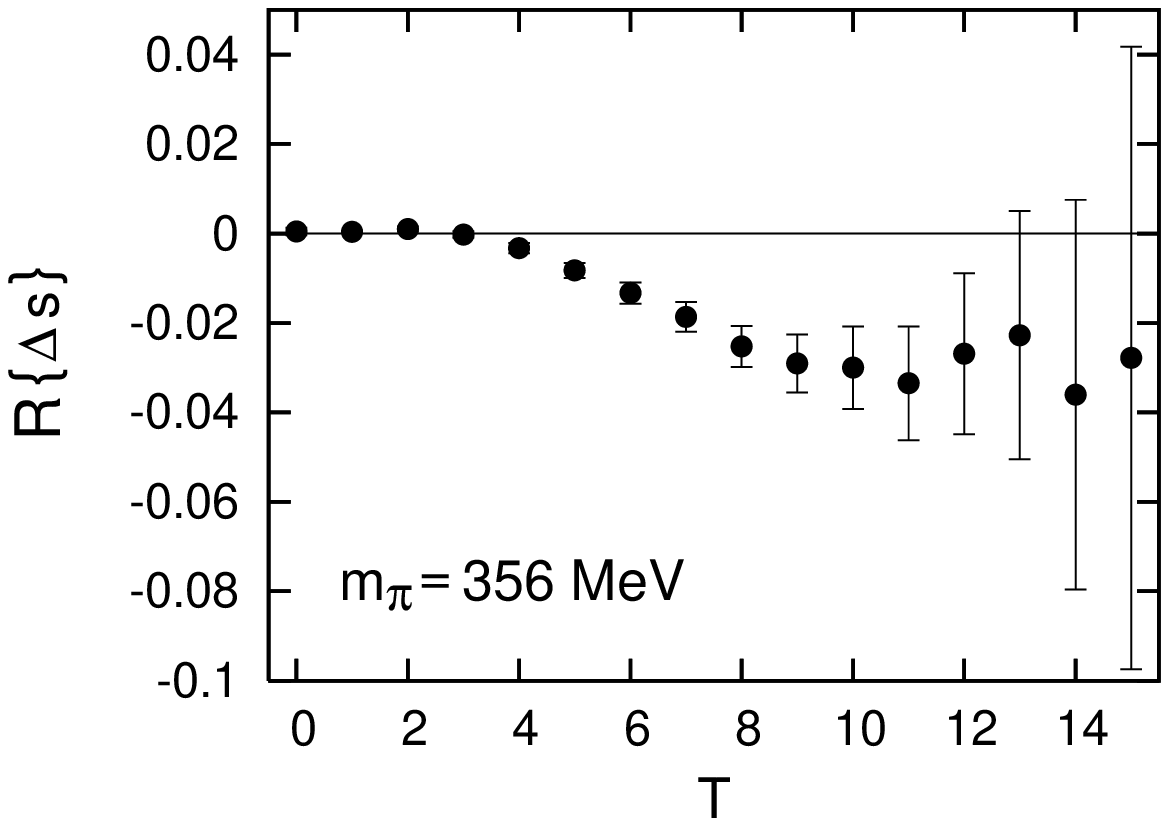} }
\vspace{0.3cm}
\centerline{\includegraphics[angle=-90,width=9cm]{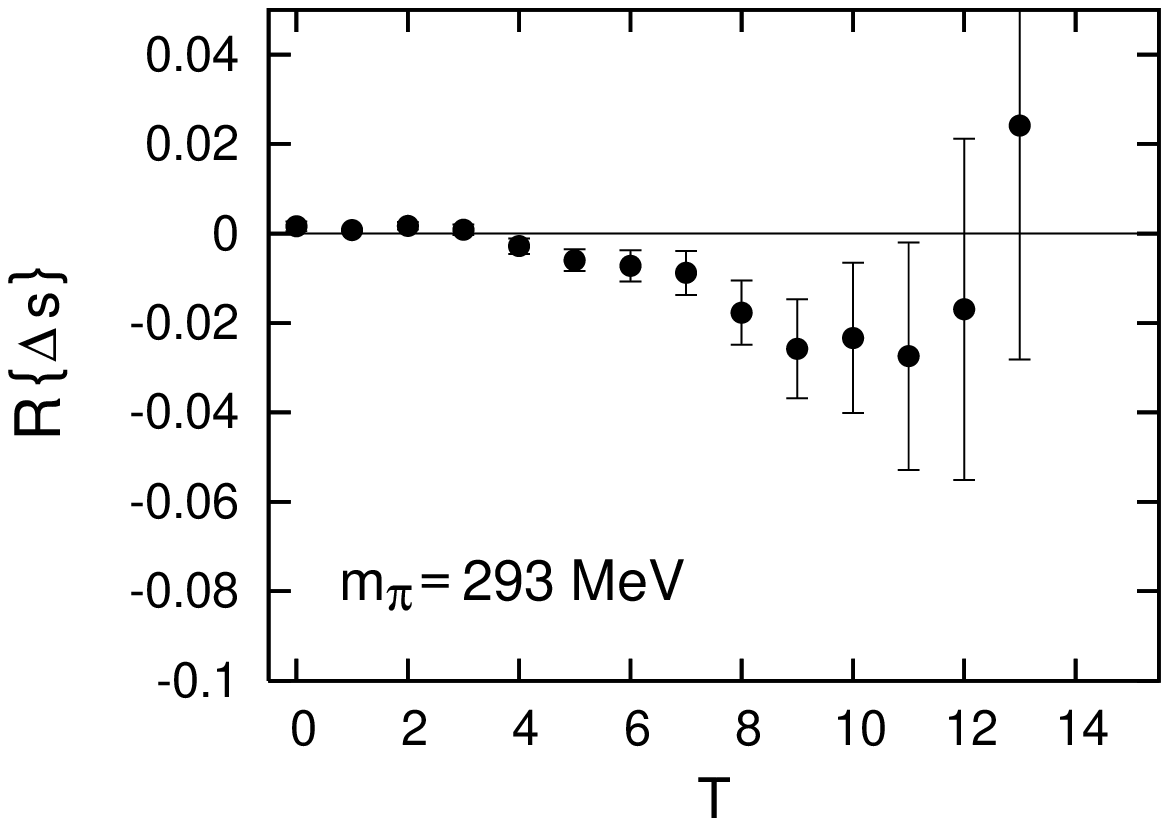} }
\vspace{0.3cm}
\caption{Correlator ratio $R\{ \Delta s \} $, cf.~(\ref{corrds}),
averaged over insertion time $\tau $ as described in section \ref{setupsec},
as a function of sink time $T$, for the three pion masses considered.}
\label{resds}
\end{figure}

The numerical results for the correlator ratios $R\{ f_{T_s } \} $ and
$R\{ \Delta s \} $, cf.~(\ref{corrfts}) and (\ref{corrds}), averaged
over the insertion time $\tau $ as described in section \ref{setupsec},
are shown in Figs.~\ref{resfts} and \ref{resds} as a function of sink
time $T$. The correlator ratios start out near vanishing values at small
sink times $T$, and then accumulate strength as $T$ traverses the region
of support of the stochastic strange quark source. Beyond this region,
$R\{ f_{T_s } \} $ and $R\{ \Delta s \} $ begin to level off and approach
their asymptotic values $f_{T_s } $ and $\Delta s $, respectively. There
is only a limited time window in which the latter behavior can be observed
due to the increase in statistical fluctuations for large $T$.

\begin{table}[h]
\begin{center}
\begin{tabular}{|c|c|c|}
\hline
$m_{\pi } $ & $R\{ \Delta s \} |_{T=10} $ &
$R\{ \Delta s \} |_{T=10,\ldots ,12} $ \\
\hline \hline
$293\, \mbox{MeV} $ & -0.023(17) & -0.023(25) \\
\hline
$356\, \mbox{MeV} $ & -0.030(9) & -0.030(12) \\
\hline
$495\, \mbox{MeV} $ & -0.021(4) & -0.022(5) \\
\hline
\end{tabular}
\end{center}
\caption{Correlator ratio $R\{ \Delta s \} $ at sink time $T=10$, and
averaged over $T=10,\ldots ,12$. In the latter case, the error estimate
is obtained by the jackknife method, i.e., the correlations between
values of $R\{ \Delta s \} $ at neighboring sink times are taken into
account.}
\label{dsplat}
\end{table}
The correlator ratio $R\{ \Delta s \} $ consistently behaves in line with
the expectation advanced in section \ref{setupsec}, namely, there is no
significant trend in the correlator ratio beyond $T=10$. This is
quantified in Table~\ref{dsplat}, which compares the value of
$R\{ \Delta s \} $ at $T=10$ with its $T=10,\ldots ,12$ average.

The behavior of the lattice data for the correlator ratio $R\{ f_{T_s } \} $
is not as smooth as in the case of $R\{ \Delta s \} $. In particular, in the
$m_{\pi } = 356\, \mbox{MeV} $ correlator ratio, one notices an enhancement
in the $T=12,\ldots ,14$ region compared with the value at $T=10$. On the
other hand, both in the $m_{\pi } = 495\, \mbox{MeV} $ and the
$m_{\pi } = 293\, \mbox{MeV} $ data, plateaux appear to be reached
at $T=10$. Table~\ref{ftsplat} again provides a comparison of the
$T=10$ value of $R\{ f_{T_s } \} $ with its $T=10,\ldots ,12$ average;
in the case of $m_{\pi } = 356\, \mbox{MeV} $, the results are barely
compatible within the statistical uncertainties. Nevertheless, an
interpretation of this behavior as a statistical fluctuation, as opposed
to a systematic effect, seems most plausible: The direction of the
deviations from the expected plateau behavior in the correlator ratio
beyond $T=10$ is not consistent across the data sets, with
$R\{ f_{T_s } \} $ decreasing slightly for $m_{\pi } = 495\, \mbox{MeV} $,
while it rises in the other two cases. Furthermore, there is no clear
trend as a function of pion mass, with the largest upward deviation
occurring for the middle pion mass, $m_{\pi } = 356\, \mbox{MeV} $, while
at $m_{\pi } = 293\, \mbox{MeV} $ the deviation is again quite small.
\begin{table}[t]
\begin{center}
\begin{tabular}{|c|c|c|}
\hline
$m_{\pi } $ & $R\{ f_{T_s } \} |_{T=10} $ &
$R\{ f_{T_s } \} |_{T=10,\ldots ,12} $ \\
\hline \hline
$293\, \mbox{MeV} $ & 0.060(10) & 0.064(14) \\
\hline
$356\, \mbox{MeV} $ & 0.043(6) & 0.051(8) \\
\hline
$495\, \mbox{MeV} $ & 0.046(3) & 0.045(4) \\
\hline
\end{tabular}
\end{center}
\caption{Correlator ratio $R\{ f_{T_s } \} $ at sink time $T=10$, and
averaged over $T=10,\ldots ,12$. In the latter case, the error estimate
is obtained by the jackknife method, i.e., the correlations between
values of $R\{ f_{T_s } \} $ at neighboring sink times are taken into
account.}
\label{ftsplat}
\end{table}

In summary, thus, at the level of statistical uncertainty achieved in the
present calculation, no systematic excited state effects stemming from
a too restricted source-sink separation can be reliably extracted; or,
in other words, such effects are smaller than the aforementioned statistical 
uncertainties. In view of this, in the following, the $T=10$ values of the
correlator ratios $R\{ \Delta s \} $ and $R\{ f_{T_s } \} $, as reported
in Tables~\ref{dsplat} and \ref{ftsplat}, will be regarded as the most
reliable estimates for their asymptotic limits $\Delta s$ and $f_{T_s } $,
respectively. Systematic uncertainties, including the ones due to
excited states, will be revisited and discussed in detail in
section~\ref{syssec}.

\section{Renormalization}
\label{renormsec}
To establish a connection with phenomenology, quantities measured on
the lattice need to be related to their counterparts in a standard
renormalization scheme such as the $\overline{MS} $ scheme at a scale
of $2\, \mbox{GeV} $. An advantage of the domain wall fermion
discretization, used in the present treatment to represent the strange
quark fields entering the quark bilinear operator insertions in the matrix
elements (\ref{matel1}) and (\ref{matel2}), lies in its good chiral symmetry
properties, which lead to a mild renormalization behavior. As a result,
even though not all elements necessary for a complete renormalization of
the quantities considered are available (e.g., matrix elements of gluonic
operators in the nucleon, with which the flavor singlet parts of the
strange quark bilinears may mix), it is possible to estimate the
associated uncertainties in the renormalization, indicating that
these do not dominate over the statistical uncertainties of the
calculation.

\subsection{Scalar matrix element}
\label{scalrensec}
Consider first the renormalization of $f_{T_s } $, specifically, the
proposition that the combination $m_s \langle N | \bar{s} s | N \rangle $
can be treated as independent of the renormalization scheme and scale,
\begin{equation}
\left. m_s \langle N | \bar{s} s | N \rangle
\right|_{\mbox{\scriptsize renorm} } \stackrel{?}{=}
\left. m_s \langle N | \bar{s} s | N \rangle
\right|_{\mbox{\scriptsize bare} } \ .
\label{scalarinv}
\end{equation}
As has been emphasized previously in \cite{jlqcd2}, this behavior is
contingent upon chiral symmetry being maintained in the lattice evaluation
of the bare quantities. In general, the strange scalar operator $\bar{s} s$
can mix both with light quark and gluonic operators; decomposing it into
flavor singlet and octet components,
\begin{equation}
\bar{s} s = \frac{1}{3} \left[ \bar{q} q - \sqrt{3} \bar{q} \lambda_{8} q
\right]
\label{singoct}
\end{equation}
(where $q\in \{ u,d,s \} $), the flavor singlet part can mix with the
relevant gluonic operator,
\begin{equation}
(\bar{q} q)^{\mbox{\scriptsize renorm} } =
Z_{S}^{00} (\bar{q} q)^{\mbox{\scriptsize bare} } +
Z_{S}^{0G} (F^2 )^{\mbox{\scriptsize bare} }
\label{gluemix}
\end{equation}
and also the renormalization constant $Z_{S}^{88} $ of the octet part,
\begin{equation}
(\bar{q} \lambda_{8} q)^{\mbox{\scriptsize renorm} } =
Z_{S}^{88} (\bar{q} \lambda_{8} q)^{\mbox{\scriptsize bare} }
\end{equation}
in general is not identical to $Z_{S}^{00} $. However, these mixing effects
would have to proceed \cite{jlqcd2} via diagrams in which the
$\bar{q} q$ insertion occurs within a disconnected quark loop; furthermore,
since $\bar{q} q = \bar{q}_{L} q_R + \bar{q}_{R} q_L $, a chirality flip
would have to take place within that loop. Gluon vertices coupling to the
loop cannot effect such a chirality flip. Thus, if all explicit chiral
symmetry breaking effects are excluded, both in the dynamics as well as
by adopting a mass-independent renormalization scheme, the gluonic
admixture in (\ref{gluemix}) is avoided. By extension
\cite{jlqcd2}, also no mixing of $\bar{s} s $ with
$\bar{u} u+\bar{d} d$ can take place, i.e., the renormalization factors
of the flavor singlet and octet parts in (\ref{singoct}) are equal under
the assumption of strict chiral symmetry. Then, $\bar{s} s $ renormalizes
in a purely multiplicative manner.

Complementarily, also the strange quark mass $m_s $ only renormalizes
multiplicatively, $m_s^{\mbox{\scriptsize renorm} } =
Z_m m_s^{\mbox{\scriptsize bare} } $, when chiral symmetry is maintained.
For generic non-chiral lattice fermion discretizations, the renormalized
strange quark mass $m_s^{\mbox{\scriptsize renorm} } $ acquires strong
additive contributions supplementing the aforementioned multiplicative
renormalization, invalidating the behavior (\ref{scalarinv}). This is
again avoided when chiral symmetry is respected. Indeed, when renormalization
is purely multiplicative, in view of the Feynman-Hellmann theorem,
\begin{equation}
m_s \langle N | \bar{s} s | N \rangle =
m_s \frac{\partial m_N }{\partial m_s } \ ,
\label{fhtheorem}
\end{equation}
factors $Z_m $ acquired by the strange quark mass under renormalization
cancel on the right-hand side of (\ref{fhtheorem}), implying that also
the left-hand side is invariant.

Chiral symmetry of the lattice fermion discretization is therefore 
instrumental in establishing a simple renormalization behavior of
$f_{T_s } $, and, indeed, is realized to a good approximation
in the present treatment due to the use of domain wall fermions in the
coupling to the operator $\bar{s} s$. Whereas exact chiral symmetry
would only be achieved using an infinite fifth dimension in the domain
wall construction, in practice, the extent of this dimension has been
chosen sufficiently large as to render the residual mass $m_{res} $
quantifying the violation of chiral symmetry an order of magnitude smaller
than the light quark mass in all ensembles considered \cite{lhpc07}.
One must, however, be careful in assessing the significance
of this smallness, since, on the other hand, the light quark and
gluonic operators with which $\bar{s} s$ mixes may have nucleon matrix
elements much larger than $\langle N | \bar{s} s | N\rangle $; for
$\langle N | \bar{u} u + \bar{d} d | N\rangle $, this is certainly
the case. This can potentially offset the suppression of $m_{res} $.
One may speculate that mixing with the operator $\bar{u} u + \bar{d} d$
constitutes the strongest effect, since the light quark fields are special
in that they form the valence component of the nucleon, which has no
counterpart in the vacuum expectation value that is subtracted off
throughout, cf.~(\ref{matel1}). On the other hand, the presence of the
valence quarks also strongly distorts the gluon field in the nucleon. No
estimate of the gluonic admixture to $\bar{s} s$ is available, but the
light quark admixture under renormalization will be argued below to
constitute an effect of the order of 1\%. In view of the statistical
uncertainties associated with the present determination of $f_{T_s } $,
which amount to about 20\%, a putative gluonic mixing effect would have to
be an order of magnitude larger than the light quark mixing effect in order
to appreciably influence the final result for $f_{T_s } $. This seems
a rather implausible scenario. For this reason, the strong constraint
on mixing with light quarks derived below will be taken as indication
that violations of (\ref{scalarinv}) are negligible at the present level
of statistical accuracy of $f_{T_s } $.

Concentrating thus on the effect stemming from the mixing with the
operator $\bar{u} u + \bar{d} d$, an estimate of the possible residual
violation of (\ref{scalarinv}) can be obtained from the Feynman-Hellmann
theorem as follows \cite{jlqcd2}. The residual breaking of chiral symmetry
can be parametrized to leading order via the additive mass renormalization
$m_{res,q} $, which in general depends on the flavor $q$ for which one is
considering the domain wall Dirac operator,
\begin{equation}
m_q^{\mbox{\scriptsize renorm} } =
Z_m (m_q^{\mbox{\scriptsize bare} } + m_{res,q} ) \ .
\end{equation}
Using (\ref{fhtheorem}), one has
\begin{eqnarray}
\left. m_s \langle N |\bar{s} s| N\rangle \right|_{\mbox{\scriptsize bare} }
&=& m_s^{\mbox{\scriptsize bare} } \sum_{q}
\frac{\partial m_N }{\partial m_q^{\mbox{\scriptsize renorm} } }
\frac{\partial
m_q^{\mbox{\scriptsize renorm} } }{\partial m_s^{\mbox{\scriptsize bare} } }
\label{scalmix} \\ & \approx &
\left. m_s \langle N |\bar{s} s| N\rangle \right|_{\mbox{\scriptsize renorm} }
+ m_s^{\mbox{\scriptsize renorm} }
\frac{\partial m_{res,l} }{\partial m_s^{\mbox{\scriptsize bare} } }
\langle N |\bar{u} u + \bar{d} d | N \rangle^{\mbox{\scriptsize renorm} }
\nonumber
\end{eqnarray}
where $m_{res,l} $ denotes the light quark flavor residual mass, and where
only the dominant correction has been kept in the second line by using
$m_{s}^{\mbox{\scriptsize bare} } \gg m_{res,s} $
and $\langle N |\bar{u} u + \bar{d} d | N\rangle \gg
\langle N |\bar{s} s | N\rangle $. No direct data for the factor
$\partial m_{res,l} /\partial m_s^{\mbox{\scriptsize bare} } $ are
available, but a rough order of magnitude estimate can be constructed from
related data on the residual mass obtained within the LHPC program and
in the present work. This estimate is discussed in the Appendix, with
the result
\begin{equation}
\left.
\frac{\partial m_{res,l} }{\partial m_s^{\mbox{\scriptsize bare} } }
\right|_{am_s =0.081, am_l =0.0081} \approx -0.00035
\end{equation}
cf.~(\ref{simpest}), at the lowest light quark mass considered in the
numerical calculations in this work. In turn, an upper bound for the
light quark scalar matrix element
$\langle N |\bar{u} u + \bar{d} d | N \rangle^{\mbox{\scriptsize renorm} } $
is given by its magnitude at the physical point; combining a typical
phenomenological value for the nucleon sigma term,
$m_l \langle N |\bar{u} u + \bar{d} d | N \rangle \approx 60\, \mbox{MeV} $,
cf., e.g., \cite{camalich} for a recent discussion, with the physical
light quark mass \cite{pdg}, $m_l \approx 3.5\, \mbox{MeV} $, yields
$\langle N |\bar{u} u + \bar{d} d | N \rangle \approx 17$. Supplementing this
with the value of the strange quark mass \cite{pdg}, $m_s =95\, \mbox{MeV} $
(all aforementioned quark masses being quoted in the $\overline{MS} $
scheme at $2\, \mbox{GeV} $), yields
\begin{equation}
\delta (m_s \langle N| \bar{s} s | N\rangle ) \approx -0.6\, \mbox{MeV}
\label{ftscorrest}
\end{equation}
for the correction term in the second line of (\ref{scalmix}). This would
amount to a 1\% upward shift of the bare result for $f_{T_s } $ as one
translates the quantity to the $\overline{MS} $ scheme at $2\, \mbox{GeV} $.
Compared to the statistical uncertainties of the present calculation, a
correction of this magnitude is negligible, and in the following,
(\ref{scalarinv}) will therefore be taken to hold for the present
calculation within its uncertainties\footnote{Note that the consideration
of the factor $\partial m_{res,l} /\partial m_s^{\mbox{\scriptsize bare} } $
given in the Appendix relies on a number of assumptions and should
be viewed as no more than a rough estimate of what is most likely an
upper bound on the magnitude of this correction factor. In view of this,
applying the correction (\ref{ftscorrest}) to the numerical data does not
seem warranted.}.

It should again be noted that the mixing with gluonic operators has not
been quantified in the above considerations. Whereas the weakness of the
mixing with light quark operators (which, after all, is mediated by the
coupling to the gluonic fields) makes it seem improbable that mixing with
gluonic operators themselves is significant compared to other uncertainties
of the calculation, explicit corroboration of this expectation would be
desirable.

\subsection{Axial vector matrix element}
Turning to the axial vector matrix element, chiral symmetry again
provides important constraints, although it cannot completely exclude
mixing effects, due to its anomalous $U_A (1)$ breaking. The domain wall
fermion discretization admits a five-dimensional partially conserved axial
vector current ${\cal A}_{\mu }^{a} $ obeying a Ward-Takahashi identity
of the form \cite{shamir,rbc1}
\begin{equation}
\Delta_{\mu } {\cal A}_{\mu }^{a} = 2mJ_5^a + 2J_{5q}^{a}
\label{wtid}
\end{equation}
The first term on the right-hand side represents the explicit breaking
of chiral symmetry by the quark masses, whereas the second term in
the flavor-octet case encodes the residual chiral symmetry breaking
present for a fifth dimension of finite extent \cite{rbc1},
$J_{5q}^{a} \approx m_{res} J_5^a $. Thus, up to these chiral symmetry
breaking effects, which will be revisited further below, the flavor octet
part of the current ${\cal A}_{\mu }^{a} $ is conserved and undergoes no
renormalization. On the other hand, in the flavor singlet case, the second
term on the right-hand side of (\ref{wtid}) in addition encodes the
coupling to the gluonic topological density which leads to the anomalous
$U_A (1)$ breaking of chiral symmetry. This opens the possibility of
operator mixing under renormalization in the flavor singlet component.
With respect to direct gluonic admixtures, the axial case differs somewhat
from the scalar case discussed further above. The continuum axial vector
current ${\cal A}_{\mu }^{\mbox{\scriptsize{(cont)} } } $ receives no direct
gluonic admixtures \cite{anom}, since the relevant gluonic operator would be
the operator $K_{\mu } $ which, upon taking the divergence, yields the gluonic
topological density, $\partial_{\mu } K_{\mu } = g^2 \widetilde{F} F$.
However, $K_{\mu } $ itself is not gauge invariant, and therefore the gauge
invariant operator ${\cal A}_{\mu }^{\mbox{\scriptsize{(cont)} } } $
receives no admixtures from $K_{\mu } $. While it is not clear
to what extent this argument is modified at finite lattice spacing, the
fact that direct gluonic admixtures to the flavor-singlet axial vector
current must vanish in the continuum limit suggests that any such
modifications would be small compared to the mixing of the strange
quark axial current with the light quark axial currents. Concentrating
thus on the latter, one can obtain a corresponding estimate of operator
mixing effects as follows. Decomposing
\begin{equation}
\bar{s} \gamma_{\mu} \gamma_{5} s =
\frac{1}{3} \left[ \bar{q} \gamma_{\mu} \gamma_{5} q -
\sqrt{3} \bar{q} \gamma_{\mu} \gamma_{5} \lambda_{8} q
\right]
\label{axoct}
\end{equation}
(where $q\in \{ u,d,s \} $), and allowing for different renormalization
constants for the singlet and octet parts,
\begin{eqnarray}
\left( \bar{s} \gamma_{\mu} \gamma_{5} s \right)^{\mbox{\scriptsize renorm} }
&=& \frac{1}{3} Z_{A}^{00} \bar{q} \gamma_{\mu} \gamma_{5} q
-\frac{\sqrt{3} }{3} Z_{A}^{88}
\bar{q} \gamma_{\mu} \gamma_{5} \lambda_{8} q \\
&=& Z_{A}^{88} \bar{s} \gamma_{\mu} \gamma_{5} s +
\frac{1}{3} \frac{Z_{A}^{00} -Z_{A}^{88} }{Z_{A}^{88} } Z_{A}^{88}
\bar{q} \gamma_{\mu} \gamma_{5} q
\label{axmix}
\end{eqnarray}
The relative strength of mixing $(Z_{A}^{00} -Z_{A}^{88} )/Z_{A}^{88} $
is not available for the specific lattice scheme used in the present
work, but has been estimated for the case of clover fermions
\cite{panag,bali2}. To the extent that this encodes the effect
of the axial anomaly as opposed to specific lattice discretization
effects, it can be taken as indicative of the strength of mixing also
in other lattice schemes such as the one used in the present investigation.
Inasfar as it is influenced by the lattice scheme, it presumably can be
taken to represent an upper bound for the strength of mixing in schemes
which better respect chiral symmetry such as the one used here (assuming
there are no accidental cancellations).
Quantitatively, for the conversion into the $\overline{MS} $ scheme at
the scale $\mu =2\, \mbox{GeV} $, one obtains in the two-flavor clover
fermion case \cite{bali2}
\begin{equation}
\frac{Z_{A}^{00} -Z_{A}^{88} }{Z_{A}^{88} } =
\frac{0.0082}{0.765} = 0.011
\end{equation}
Correcting by a factor $3/2$ to translate to the three-flavor case
considered here, and supplementing with $Z_{A}^{88} \approx 1.1$,
as obtained below, as well as $\Delta (u+d) \approx 0.42$ (at the
physical pion mass, cf.~\cite{lhpc10}), the uncertainty from mixing
with light quarks, i.e., the second term in (\ref{axmix}) is estimated
to amount to
\begin{equation}
\delta (\Delta s) \approx 0.0025
\label{axuncer}
\end{equation}
directly at the physical pion mass. Due to the indirect nature of this
estimate, the shift (\ref{axuncer}) will not be applied to the
(chirally extrapolated) central value for $\Delta s$ obtained in this
work, but will be treated as a systematic uncertainty.

It should be remarked that the considerations for the clover case
\cite{bali2} referred to above pertain to lattices with a
substantially finer spacing than used here ($0.073 \, \mbox{fm} $
vs.~$0.124 \, \mbox{fm} $). However, adjusting for this is expected
to modify (\ref{axuncer}) merely by a few percent and thus is negligible
in the present context. This can be inferred from the magnitude of
the $O(a)$ improvement corrections to the renormalization constants
quoted in \cite{bali2}; note also that the present HYP-smeared mixed
action scheme, which is fully $O(a)$-improved, suffers only from very
benign finite lattice spacing effects, as evidenced, e.g., by the
congruence between nucleon mass measurements in this same scheme and
corresponding MILC determinations on much finer, $a=0.06\, \mbox{fm} $
lattices \cite{lhpc08}.

Turning to the renormalization constant $Z_{A}^{88} $, the axial vector
quark bilinear used in practice in evaluating $\Delta s$ is the local
$\bar{s} (x) \gamma_{\mu } \gamma_{5} s(x)$ as opposed to the corresponding
flavor component of the partially conserved ${\cal A}_{\mu } $ of
eq.~(\ref{wtid}). To renormalize this local operator, the standard
scheme \cite{rbc1} can be applied, with the modification that, in the
present case, it is not the pion current but the $\eta $ current which
is relevant: Calculating the (connected parts of the) two-point functions
of both the conserved and the local current,
\begin{eqnarray}
C(t+1/2) &=& \sum_{x} \langle {\cal A}_{0}^{8} (x,t) \ \eta (0,0)
\rangle |_{\mbox{\scriptsize conn} } \nonumber \\
L(t) &=& \sum_{x} \langle \bar{q} \gamma_{0} \gamma_{5} (\lambda_{8} /2) q \
\eta (0,0) \rangle |_{\mbox{\scriptsize conn} } \ ,
\label{twopteta}
\end{eqnarray}
$Z^{88}_{A} /Z^{88}_{\cal A} $ can be extracted from an appropriate ratio
which takes into account the temporal offset between the two currents,
\begin{equation}
\frac{1}{2} \left( \frac{C(t+1/2) + C(t-1/2)}{2L(t)}
+\frac{2C(t+1/2)}{L(t)+L(t+1)} \right)
\stackrel{t/a \gg 1}{\longrightarrow } \frac{Z^{88}_{A} }{Z^{88}_{\cal A} }
\end{equation}
Note that the full correlators $\langle \bar{q} \gamma_{0} \gamma_{5}
(\lambda_{8} /2) q \ \eta (0,0) \rangle $ and
$\langle {\cal A}_{0}^{8} (x,t) \ \eta (0,0) \rangle $ acquire
disconnected contributions for unequal strange and light quark
masses; however, for the specific purpose of extracting
$Z^{88}_{A} /Z^{88}_{\cal A} $, any ratio of quantities which only
differ by this overall renormalization factor is suitable, including using
only the connected parts of the aforementioned correlators, as indicated
in (\ref{twopteta}). Table~\ref{zatab} lists the values obtained for
$Z^{88}_{A} /Z^{88}_{\cal A} $, which are applied to the bare lattice
measurements of $\Delta s$.
\begin{table}[h]
\begin{center}
\begin{tabular}{|c|c|c|c|c|}
\hline
$m_l^{\mbox{\scriptsize bare} } $ & 0.0081 & 0.0138 & 0.0313 & 0.081 \\
\hline
$Z^{88}_{A} /Z^{88}_{\cal A} $ & 1.09 & 1.09 & 1.10 & 1.13 \\
\hline
\end{tabular}
\end{center}
\caption{Renormalization factor $Z^{88}_{A} /Z^{88}_{\cal A} $ at
varying $m_l^{\mbox{\scriptsize bare} } $.}
\label{zatab}
\end{table}

A systematic uncertainty is associated with this scheme of renormalizing
$\Delta s$, due to residual sources of chiral symmetry breaking. One
of these sources is the finite extent of the fifth dimension in the
domain wall fermion construction. At finite $m_{res} $, the
renormalization constant of the partially conserved current,
$Z^{88}_{\cal A} $, can deviate from the unit value it would take if
chiral symmetry were strictly observed \cite{rbc2}. Secondly,
note that, at finite lattice spacing $a$, there is a certain
tension between adopting a mass-independent lattice renormalization scheme
and maintaining $O(a)$ improvement \cite{luescher}. The renormalization
constant $Z^{88}_{A} $ in general contains dependences of order
$O(m_q a)$, evident in the slight variation displayed in
Table~\ref{zatab}. Since the lattice data necessary to perform the
continuum limit $a\rightarrow 0$ are not available, two options remain:
One option would be to extrapolate $Z^{88}_{A} /Z^{88}_{\cal A} $ to the
chiral limit, thus obtaining a mass-independent renormalization scheme in
more direct correspondence to the $\overline{MS} $ scheme, but spoiling
$O(a)$ improvement. On the other hand, by retaining the leading quark
mass dependence, i.e., applying the finite-$m_{q} $ renormalization
constants in Table~\ref{zatab} ensemble by ensemble, $O(a)$ improvement is
maintained, at the expense of introducing a slight mass dependence into
the renormalization scheme at finite $a$. The mass dependence, implying
a breaking of chiral symmetry in addition to the one encoded in the
residual mass $m_{res} $, is then expected to be of order $O(m_q a^2 )$.
Since the present investigation yielded lattice data at only a single
lattice spacing $a$, precluding a direct estimate of the effects of
finite lattice spacing, maintaining $O(a)$ improvement seems sufficiently
desirable to elect the latter alternative, i.e., applying the
renormalization constants in Table~\ref{zatab} ensemble by ensemble
and thus introducing a slight mass dependence into the renormalization
scheme.

A way of estimating the systematic uncertainty in the renormalization
of $\Delta s$ resulting from the residual breaking of chiral symmetry
due to the above sources lies in the mismatch between the axial vector
and the vector renormalization factors, $Z_A /Z_{\cal A} $
vs.~$Z_V /Z_{\cal V} $, which would remain equal if chiral symmetry
were strictly maintained \cite{rbc3}. In the lattice scheme used here,
these factors typically differ by $3\% $ \cite{lhpc10}\footnote{To be
precise, \cite{lhpc10} considered the isovector currents, i.e.,
determined $Z_A^{33} /Z_{\cal A}^{33} $ and $Z_V^{33} /Z_{\cal V}^{33} $. }.
An additional systematic uncertainty of this magnitude will therefore be
attached to the renormalized value of $\Delta s$.

\section{Chiral extrapolation}
\label{chisec}
Chiral extrapolation formulae for strange quark matrix elements in the
nucleon have been given in \cite{chesav}. At leading order (LO), both
$\langle N | \bar{s} s | N\rangle $ and $\Delta s$ are constant in
$m_{\pi }^{2} $. On the other hand, when one evaluates one-loop effects,
including $\Delta $-resonance degrees of freedom, one obtains
next-to-next-to-leading-order (NNLO) formulae which contain too
many parameters to be effectively constrained by the restricted set
of lattice data at three pion masses accessed in this work. However,
practicable fits can be constructed by reducing the NNLO formulae
given in \cite{chesav} to the chiral effective theory without
$\Delta $-resonance degrees of freedom, which is achieved by setting
$g_{\Delta N} =0$ and correspondingly also eliminating the counterterms
associated with the $\Delta $-resonance degrees of freedom.
In this case, the behavior of $\langle N | \bar{s} s | N\rangle $
reduces to a linear function in the light quark mass, i.e., in $m_{\pi }^2 $,
\begin{equation}
\langle N | \bar{s} s | N\rangle = S_1 + S_2 m_{\pi }^{2}
\end{equation}
with the two fit parameters $S_1 $ and $S_2 $, whereas $\Delta s$ retains a
chiral logarithm,
\begin{equation}
\Delta s = D_1 \left[ 1-\frac{3g_A^2 }{8\pi^{2} f^{2} }
m_{\pi }^{2} \log (m_{\pi }^{2} /\mu^{2} ) \right] + D_2 m_{\pi }^{2}
\end{equation}
with the two fit parameters $D_1 $ and $D_2 $; the dependence on the
scale $\mu $ is of course absorbed by the $D_2 $ counterterm. The pion
decay constant $f$ is normalized such that $f\sim 132\, \mbox{MeV} $,
and the physical axial coupling constant is $g_A \sim 1.26$.
\begin{figure}
\centerline{\includegraphics[angle=-90,width=9cm]{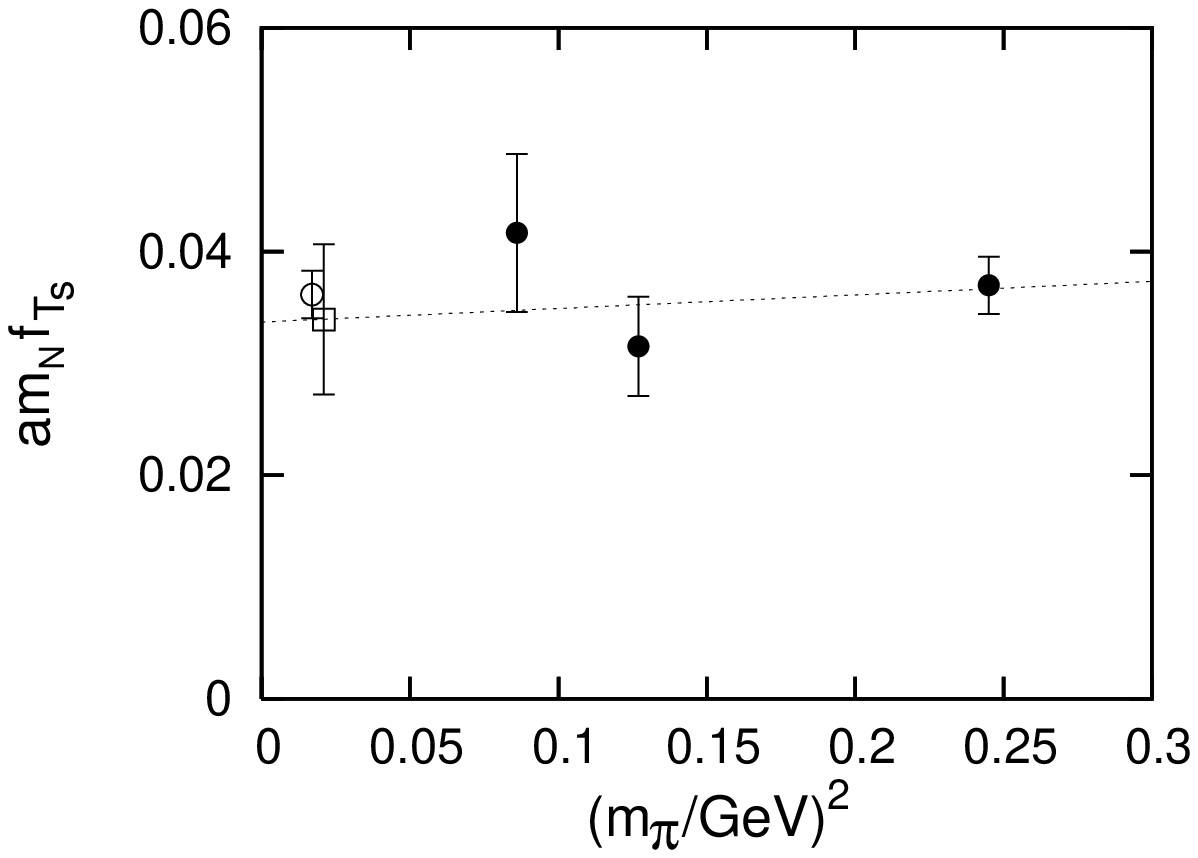} }
\vspace{0.3cm}
\centerline{\includegraphics[angle=-90,width=9cm]{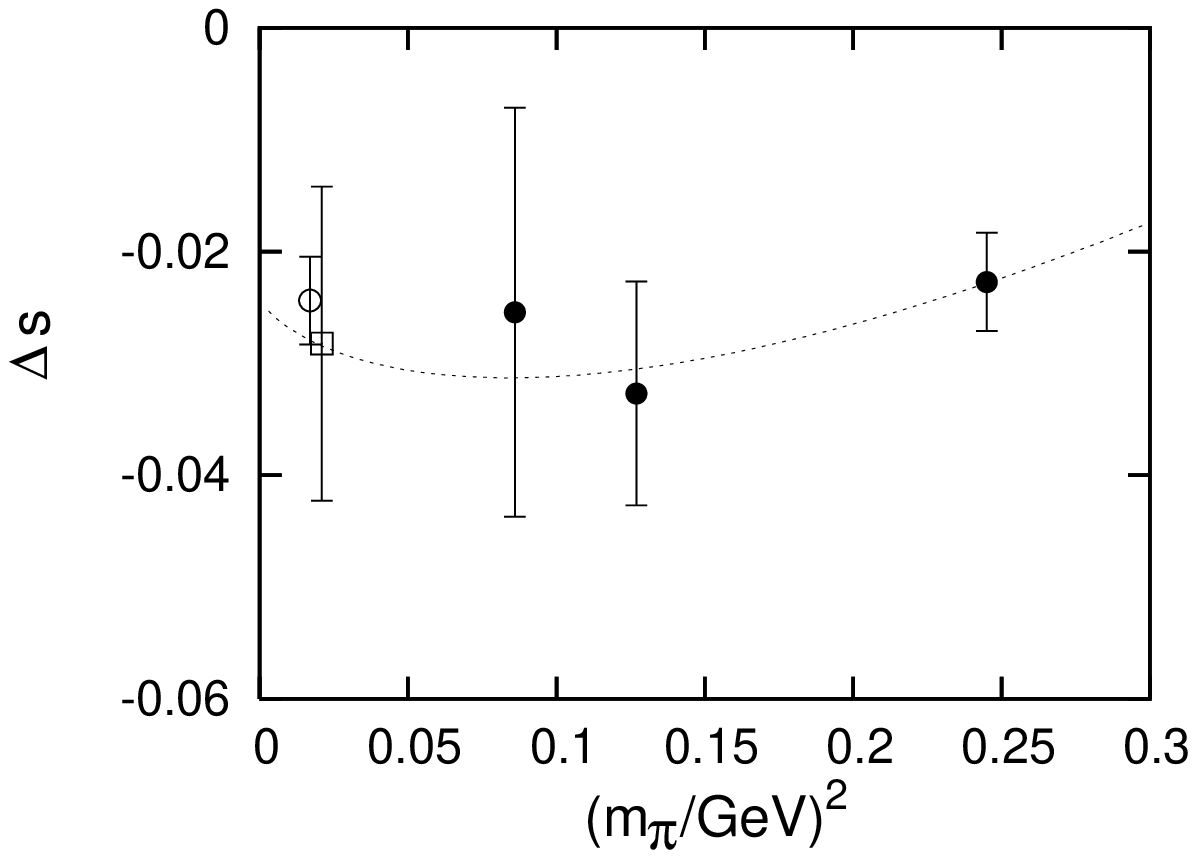} }
\vspace{0.3cm}
\caption{Pion mass dependence of the results for
$m_s \langle N| \bar{s} s |N \rangle = m_N f_{T_s } $ and $\Delta s$.
Filled circles represent renormalized lattice data; in the case of
$\Delta s$, these are obtained by multiplying the $T=10$ values from
Table~\ref{dsplat} by the corresponding renormalization constants from
Table~\ref{zatab}, whereas $m_N f_{T_s } $ is obtained by multiplying the
$T=10$ values from Table~\ref{ftsplat} by the corresponding nucleon masses
from Table~\ref{tabmilc}. Open symbols show chiral extrapolations of the
lattice data to the physical pion mass, cf.~main text. Open circles
represent the LO (constant) chiral extrapolations, whereas open squares
represent the reduced NNLO extrapolations obtained by dropping the
$\Delta $-resonance degrees of freedom, with the dashed lines showing
the pion mass dependences of the central values in the latter case.}
\label{extrap}
\end{figure}
Corresponding chiral fits to the renormalized lattice data for 
$m_s \langle N| \bar{s} s | N\rangle = m_N f_{T_s } $ and $\Delta s$ are
shown in Fig.~\ref{extrap}. The LO constant fits and the (reduced) NNLO
fits are consistent with one another. Positing the LO constant behavior in
$m_{\pi }^2 $ leads to artificially low estimates of the uncertainties; in
this case, the statistical error bars are dominated by the most accurately
determined $m_{\pi } = 495\, \mbox{MeV} $ data points, at which, on the
other hand, the chiral extrapolation formulae are the least trustworthy.
Plausible estimates of the statistical uncertainties of the extrapolated
values are given by the (reduced) NNLO fits, which allow for variation
of the observables with $m_{\pi }^2 $. The estimates of $f_{T_s } $ and
$\Delta s$ resulting from the (reduced) NNLO fits at the physical pion
mass in the $\overline{MS} $ scheme at a scale of $2\, \mbox{GeV} $ are
(before taking into account systematic effects, which are discussed in the
next section),
\begin{equation}
f_{T_s } = 0.057(11) \ ,
\label{ftsstat}
\end{equation}
where the physical nucleon mass has been used to convert the fit result
from Fig.~\ref{extrap} back to $f_{T_s } $, and
\begin{equation}
\Delta s = -0.028(14) \ .
\label{dsstat}
\end{equation}
Note again that the quoted uncertainties at this point contain only the
statistical error from the lattice measurement, propagated through
the chiral extrapolation; systematic uncertainties and adjustments
of the results (\ref{ftsstat}) and (\ref{dsstat}) are elaborated upon
in the next section.

\section{Systematic corrections and uncertainties}
\label{syssec}
Several sources of systematic uncertainty should be taken into account
with respect to the two results (\ref{ftsstat}) and (\ref{dsstat}):
\vspace{0.3cm}

\noindent
{\bf Renormalization uncertainties:} Uncertainties associated with
renormalization were already discussed in section~\ref{renormsec}. In the
case of $f_{T_s } $, uncertainties due to mixing with light quark operators
generated by residual breaking of chiral symmetry were estimated to be
of the order of 1\% and will thus not be considered further here. In the
case of $\Delta s$, mixing effects were less well constrained because
of the anomalous breaking of chiral symmetry. The potential correction
to $\Delta s$ was estimated to amount to $\delta (\Delta s) \approx 0.0025$
towards less negative values, cf.~(\ref{axuncer}). In addition,
a 3\% uncertainty was assigned to the renormalization factor
$Z_{A}^{88} $ due to residual sources of chiral symmetry breaking.
These included finite-$m_{res} $ effects as well as effects of
adopting a not fully mass-independent lattice renormalization scheme,
in order to preserve $O(a)$ improvement. Adding the uncertainty from
operator mixing and the one associated with $Z_{A}^{88} $ in quadrature
implies an uncertainty in $\Delta s$ of $+0.003/-0.001$.
\vspace{0.3cm}

\noindent
{\bf Finite lattice spacing effects:}
Since the present investigation only employed ensembles at a single lattice
spacing, $a=0.124\, \mbox{fm} $, no direct assessment of the $a$-dependence
of the results was possible. This motivated the insistence on a fully
$O(a)$-improved calculational scheme, in order to minimize the influence
of the finite lattice spacing from the outset (and, in the process, part
of the lattice spacing dependence was already subsumed under the
uncertainty in renormalization, as noted above). In the case of $f_{T_s } $,
which is related via the Feynman-Hellmann theorem to the nucleon mass, an
estimate of the uncertainty due to discretization effects can be inferred
from the $a$-dependence of the nucleon mass. As shown in \cite{lhpc08},
already at $a=0.124\, \mbox{fm} $, the nucleon mass in the present
HYP-smeared mixed action scheme coincides with the MILC $a=0.06\, \mbox{fm} $
results. Given that the MILC results themselves still change by about 10\%
going from $a=0.124\, \mbox{fm} $ to $a=0.06\, \mbox{fm} $, the residual
$O(a^2 )$ effect in the latter case is expected to be around 3\%. This
will therefore be taken as the generic estimate of the magnitude of finite
lattice spacing effects in the present calculation, both for $f_{T_s } $
and $\Delta s$.
\vspace{0.3cm}

\noindent
{\bf Uncertainty due to truncation of the chiral perturbation series:}
Taking the deviation between the LO and the (reduced) NNLO fits at the
physical pion mass, cf.~Fig.~\ref{extrap}, as a measure of the uncertainty
due to truncation of the chiral perturbation series, a 6\% uncertainty is
attached to the value of $f_{T_s } $ and a 14\% uncertainty to $\Delta s$.
\vspace{0.3cm}

\noindent
{\bf Effects of inadequate lattice dimensions:} Both the spatial extent
of the lattice and the temporal separations between nucleon source, sink
and operator insertion are limited. Consequently, results are influenced
both by interactions with periodic copies of the lattice as well as
excited state admixtures. Neither of the corresponding uncertainties
were directly quantifiable within the present calculation. On the one
hand, ensembles with only a single lattice extent were employed; on the
other hand, no systematic excited state effects could be gleaned from the
sink position dependence of the lattice data at the present level of
statistical accuracy, as discussed in section~\ref{resultsec}. An
indication of the possible magnitude of such effects can be inferred
from lattice calculations of the nucleon axial charge $g_A $, the
isovector light quark analogue of $\Delta s$, which represents a
well-studied benchmark quantity. Calculations of $g_A $ within the
present lattice scheme \cite{lhpc10} and others \cite{rbc4,physpi}
exhibit a deviation from the phenomenological expectation of up to
10\%, with the cause of this deviation attributed to either excited
state contaminations or finite lattice size effects. This will
therefore be taken as an estimate of the systematic uncertainty due
to such effects also in the present calculation.
\vspace{0.3cm}

\noindent
{\bf Adjustment of the strange quark mass:} The strange quark mass
$am_s^{asq} =0.05$ in the gauge ensembles used in the present calculation
lies appreciably above the physical strange quark mass, which a posteriori
was determined to be $am_s^{asq,phys} =0.036$ \cite{freeman1}. For the
case of the strange scalar matrix element, a corresponding correction
factor was estimated in \cite{freeman1}, namely,
\begin{equation}
\frac{\partial }{\partial m_s^{asq} }
\frac{\partial m_N }{\partial m_s^{asq} } = -2.2\cdot 0.31 \, \mbox{fm}
= -0.68 \, \mbox{fm}
\label{mscorr}
\end{equation}
Multiplying this by the shift in $m_s^{asq} $,
$\delta m_s^{asq} = (0.036-0.05)/a$, yields, in view of the Feynman-Hellmann
theorem, an enhancement of the strange scalar matrix element by
\begin{equation}
\delta (\langle N | \bar{s} s |N\rangle )^{asq} =
\delta m_s^{asq} \frac{\partial }{\partial m_s^{asq} }
\frac{\partial m_N }{\partial m_s^{asq} } = 0.077
\label{asqshift}
\end{equation}
To translate this to the present scheme, one needs to rescale the strange
quark mass, $m_s^{asq} = (m^{asq} /m^{DWF} ) m_s^{DWF} $. The rescaling
factor $m^{asq} /m^{DWF} $ varies only weakly between $am_s^{asq} =0.05$
and $am_s^{asq} =0.02$, namely, from $0.617$ in the former case to
$0.639$ in the latter, cf.~Table~\ref{tabmilc}. Interpolating linearly,
the most appropriate value for the present consideration is the one
halfway between $am_s^{asq} =0.05$ and $am_s^{asq} =0.036$, i.e.,
$m^{asq} /m^{DWF} = 0.622$. Since (\ref{asqshift}) has two powers of
$m_s^{asq} $ in the denominator and one in the numerator, altogether,
translated to the present scheme,
\begin{equation}
\delta (\langle N | \bar{s} s |N\rangle )^{DWF} =
\frac{m^{asq} }{m^{DWF} } \delta (\langle N | \bar{s} s |N\rangle )^{asq}
=0.048
\label{dwfshift}
\end{equation}
To obtain a measure of the relative change in
$\langle N | \bar{s} s |N\rangle $ implied by this, note that the
result (\ref{ftsstat}), multiplied by $m_N /m_s^{DWF} =7.3$, yields
$\langle N | \bar{s} s |N\rangle^{DWF}_{am_s^{DWF} =0.081} =0.42$,
and therefore the adjustment (\ref{dwfshift}) corresponds to an
enhancement of $\langle N | \bar{s} s |N\rangle $ by a factor $1.115$
as one shifts the strange quark mass to its physical value.

On the other hand, $f_{T_s } $ itself acquires an additional factor 
$m_s^{DWF,phys} /m_s^{DWF} \approx m_s^{asq,phys} /m_s^{asq} = 0.72$
as one shifts the strange quark mass to the physical point, implying
that $f_{T_s } $ is reduced by a factor $0.72\cdot 1.115 =0.80$. The
reduction of $m_s $ in fact overcompensates the enhancement of
$\langle N | \bar{s} s |N\rangle $. Altogether, thus, (\ref{ftsstat})
will be adjusted by a factor $0.80$ to arrive at the physical value;
in addition, a systematic uncertainty of 3\% will be associated with
that adjustment in view of a 15\% uncertainty in (\ref{mscorr}),
cf.~\cite{freeman1}, as well as the variability in $m^{asq} /m^{DWF} $.

For the case of $\Delta s$, no similarly detailed consideration is
available. However, it seems plausible that the leading effect of a
lowering of the strange quark mass is an overall enhancement of the
strange quark density, with the detailed dynamics governing any given
strange quark unaffected to a first approximation. In this case, one
would expect the enhancement factor 1.115 to equally apply to the
axial matrix element $\Delta s$. In view of the rough nature of this
argument, a systematic uncertainty of 12\% will be associated with this
enhancement, thus covering the range of no enhancement of $\Delta s$
up to twice the enhancement seen in the case of the scalar matrix
element.
\vspace{0.3cm}

Summarizing the diverse uncertainties and adjustments discussed above,
the final estimates for the physical values of $f_{T_s } $ and $\Delta s$
are as follows:
\begin{equation}
f_{T_s } = 0.046(9)(1)(3)(5)(1)
\end{equation}
where the uncertainties are, in the order written, statistical,
due to finite lattice spacing, due to truncation of the chiral
perturbation series, due to inadequate lattice dimensions, and
due to the adjustment of the strange quark mass to the physical
value. In turn,
\begin{equation}
\Delta s = -0.031(16)(^{+3}_{-1})(1)(4)(3)(4)
\end{equation}
where the uncertainties are, in the order written, statistical,
due to renormalization, due to finite lattice spacing, due to truncation
of the chiral perturbation series, due to inadequate lattice dimensions,
and due to the adjustment of the strange quark mass to the physical
value. To quote a succinct final result, if one combines all the systematic
uncertainties discussed in this section together with the statistical
uncertainty in quadrature,
\begin{eqnarray}
f_{T_s } &=& 0.046(11) \label{ftsfinal} \\
\Delta s &=& -0.031(17) \label{dsfinal}
\end{eqnarray}

\section{Conclusions}
\label{concsec}
This investigation focused on two of the most basic signatures of strange
quark degrees of freedom in the nucleon, namely, the strange quark
contribution to the nucleon mass, characterized by $f_{T_s } $, and the
portion of the nucleon spin contained in strange quark spin, $\Delta s$.
A high amount of averaging not only of the disconnected strange quark
loop, but also of the nucleon two-point function led to clear signals
for both $f_{T_s } $ and $\Delta s$ especially at the heaviest pion
mass, $m_{\pi } = 495\, \mbox{MeV} $, with the signals deteriorating,
but not disappearing, as the pion mass is lowered to
$m_{\pi } = 356\, \mbox{MeV} $ and $m_{\pi } = 293\, \mbox{MeV} $.
Combining all the lattice data, the signals survive chiral extrapolation to
the physical pion mass. Systematic uncertainties remain under adequate
control; the only source of systematic uncertainty which was not
quantified is gluonic operator admixtures to $f_{T_s } $ under
renormalization. However, as discussed in section~\ref{scalrensec},
a scenario in which these admixtures rise to the level at which they
begin to appreciably influence the conclusions reached regarding
$f_{T_s } $ seems highly implausible. The gluonic admixtures would have
to be an order of magnitude larger than the related light quark admixtures,
which were constrained to the 1\% level. Nevertheless, a more quantitative
corroboration of this argument would be desirable. All other systematic
uncertainties were quantified, cf.~section~\ref{syssec}, and, while some
of them are still sizeable, none rise to the level of the statistical
uncertainties. With respect to controlling systematic uncertainties,
the use of a (to a very good approximation) chirally symmetric
discretization of the strange quark fields in the matrix elements
(\ref{matel1}),(\ref{matel2}) proved very advantageous, since it provides
for benign renormalization properties, including the almost complete
suppression of light quark admixtures to $f_{T_s } $ alluded to above.
This stands in contrast to, e.g., the case of Wilson fermions, in which
the evaluation of $f_{T_s } $ is considerably complicated by the presence
of strong additive mass renormalizations \cite{bali3,babich}.

The magnitudes obtained for both $f_{T_s } $ and $\Delta s$ appear
natural. Neither quantity is abnormally enhanced; strange quarks contribute
about 4.5\% of the nucleon mass, and the magnitude of the strange quark
spin, which is polarized opposite to the nucleon spin, amounts
to about 3\% of the latter. The conditions provided by nucleon structure
for dark matter detection via coupling of the Higgs field specifically to
the strange quark component thus do not appear to be as favorable as assumed
in the most optimistic scenaria. There is also no indication from the
result for $\Delta s$ that an unnaturally large contribution to the spin
of the nucleon is hidden in the small-$x$ strange quark sector that has
hitherto eluded experimental study. The strange quark spin does
indeed appear to be polarized slightly in the direction opposite to the
nucleon spin, as also indicated by the preponderance of phenomenological
studies. However, the magnitude of $\Delta s$ found in the present
calculation is smaller than the magnitudes extracted in the analyses
\cite{herm1,compass2} which assume a substantial enhancement of
the strange quark helicity distribution at small momentum fraction $x$.

\section*{Acknowledgments}
Fruitful exchanges with W.~Freeman, H.~Grie\ss hammer, P.~H\"agler,
K.~Orginos, S.~Pate and D.~Toussaint are gratefully acknowledged.
The computations required for this investigation were carried out at the
Encanto computing facility operated by NMCAC, using the Chroma software
suite \cite{chroma} and gauge ensembles provided by the MILC Collaboration.
This work was supported by the U.S.~DOE under grant DE-FG02-96ER40965.

\section*{Appendix: Estimate of
$\partial m_{res,l} /\partial m_s^{\mbox{\scriptsize bare} } $ }
The quantity $\partial m_{res,l} /\partial m_s^{\mbox{\scriptsize bare} } $
enters the estimate of operator mixing effects in $f_{T_s } $,
cf.~section~\ref{scalrensec}. No direct data for this quantity are
available, but an order of magnitude estimate can be constructed from
related data on the residual mass obtained within the LHPC program and
in the present work, summarized in Table~\ref{mrestab}.
\begin{table}[h]
\vspace{0.5cm}
\begin{center}
\begin{tabular}{|c||c|c|c|}
\hline
$am_l^{\mbox{\scriptsize bare} } $ & 0.0081 & 0.0313 & 0.081 \\
\hline\hline
$10^3 \cdot am_{res,l} $ & 1.6 & 1.2 & 0.7 \\
\hline
$10^4 \cdot am_{res,s} $ & 9.0 & 8.1 & 7.1 \\
\hline
\end{tabular}
\end{center}
\caption{Residual masses $m_{res,l} $ and $m_{res,s} $ at varying
$m_l^{\mbox{\scriptsize bare} } $.}
\label{mrestab}
\end{table}
It should be noted that all these results were obtained at a constant
lattice spacing $a$. Fitting parabolae to the data in Table~\ref{mrestab}
yields the following derivatives at the $SU(3)$-flavor symmetric point
$am_s = am_l = 0.081$ and at the lightest $m_l $ considered in this work,
$am_l = 0.0081$:
\begin{eqnarray}
\left.
\frac{\partial m_{res,l} }{\partial m_l^{\mbox{\scriptsize bare} } }
\right|_{am_s = am_l = 0.081} = -0.005
& \ \ &
\left.
\frac{\partial m_{res,l} }{\partial m_l^{\mbox{\scriptsize bare} } }
\right|_{am_s =0.081, am_l =0.0081} = -0.020
\label{llderivs} \\
\left.
\frac{\partial m_{res,s} }{\partial m_l^{\mbox{\scriptsize bare} } }
\right|_{am_s = am_l = 0.081} = -0.0007
& \ \ &
\left.
\frac{\partial m_{res,s} }{\partial m_l^{\mbox{\scriptsize bare} } }
\right|_{am_s =0.081, am_l =0.0081} = -0.0045
\label{slderivs}
\end{eqnarray}
The simplest estimate for 
$\partial m_{res,l} /\partial m_s^{\mbox{\scriptsize bare} } $ from this
can be obtained by noting that, at the $SU(3)$-flavor symmetric point,
\begin{equation}
\left.
\frac{\partial m_{res,l} }{\partial m_s^{\mbox{\scriptsize bare} } }
\right|_{am_s = am_l = 0.081} = \frac{1}{2}
\left.
\frac{\partial m_{res,s} }{\partial m_l^{\mbox{\scriptsize bare} } }
\right|_{am_s = am_l = 0.081} = -0.00035
\end{equation}
(where the factor 1/2 stems from the fact that
$\partial /\partial m_l = \partial /\partial m_u + \partial /\partial m_d $).
Assuming that the derivative of $m_{res,l} $ in the $m_s $-direction varies
only weakly as one changes $m_l $, one arrives at the estimate that also
\begin{equation}
\left.
\frac{\partial m_{res,l} }{\partial m_s^{\mbox{\scriptsize bare} } }
\right|_{am_s =0.081, am_l =0.0081} \approx -0.00035
\label{simpest}
\end{equation}
Note that this is most likely an upper bound in magnitude, since one would
expect the characteristics of the gauge fields entering the Dirac operator
to be dominated by the light quark degrees of freedom relative to the
strange quark degrees of freedom as $m_l $ becomes smaller.

A check on this estimate can be constructed by the following alternative
chain of reasoning. First, note that the quantity
$\partial m_{res,l} / \partial m_l^{\mbox{\scriptsize bare} } $ contains
two contributions, namely, one from the explicit variation of $m_l $ in the
Dirac operator which $m_{res,l} $ characterizes, and the other from the
implicit dependence of the gauge field ensemble on $m_l $. To estimate
$\partial m_{res,l} / \partial m_s^{\mbox{\scriptsize bare} } $, one
therefore needs to apply two correction factors to
$\partial m_{res,l} / \partial m_l^{\mbox{\scriptsize bare} } $: A factor
characterizing the proportion of the variation of $m_{res,l} $ due
specifically to the implicit variation of the gauge fields, and a factor
characterizing the strength of that variation with $m_s $ as opposed
to $m_l $. The order of magnitude of these correction factors can be
inferred as follows. The former factor is available at the $SU(3)$-flavor
symmetric point as the ratio
\begin{equation}
\left.
\frac{\partial m_{res,s} / \partial m_l^{\mbox{\scriptsize bare} } }{\partial
m_{res,l} / \partial m_l^{\mbox{\scriptsize bare} } }
\right|_{am_s = am_l = 0.081} = 0.14 \ .
\label{implicvsfull}
\end{equation}
For the purposes of the present argument, it will be assumed that this
factor only varies mildly as $m_l^{\mbox{\scriptsize bare} } $ is lowered
to $am_l^{\mbox{\scriptsize bare} } =0.0081$. On the other hand, assume
also that
\begin{equation}
\left.
\frac{\partial m_{res,s} }{\partial m_s^{\mbox{\scriptsize bare} } }
\right|_{\mbox{\scriptsize implicit} , am_l = 0.081} \approx
\left.
\frac{\partial m_{res,s} }{\partial m_s^{\mbox{\scriptsize bare} } }
\right|_{\mbox{\scriptsize implicit} , am_l = 0.0081}
\end{equation}
i.e., the implicit variation of $m_{res,s} $ via the dependence of the
gauge fields on $m_s $ changes only mildly as a function of $m_l $. Note
that, while this assumption is analogous to the one leading to
(\ref{simpest}), it is better founded since $m_{res,s} $ itself varies
less with $m_l $ than $m_{res,l} $. Again, one would expect the left-hand
side to represent an upper bound for the right-hand side. Noting that the
left-hand side is identical to one-half the quantities in the left-hand
identity in (\ref{slderivs}), one thus has
\begin{eqnarray}
0.08 = \frac{-0.0007/2}{-0.0045} &=&
\frac{\partial m_{res,s} / \partial m_s^{\mbox{\scriptsize bare} }
|_{{\mbox{\scriptsize implicit} , am_l = 0.081} } }{\partial m_{res,s} /
\partial m_l^{\mbox{\scriptsize bare} }
|_{am_l = 0.0081} } \\
& \approx & \left.
\frac{\partial m_{res,s} / \partial m_s^{\mbox{\scriptsize bare} }
|_{{\mbox{\scriptsize implicit} } } }{\partial m_{res,s} /
\partial m_l^{\mbox{\scriptsize bare} } } \right|_{am_l = 0.0081}
\label{substsl} \\
& \approx & \left.
\frac{\partial m_{res,l} / \partial m_s^{\mbox{\scriptsize bare} } }{\partial
m_{res,l} / \partial m_l^{\mbox{\scriptsize bare} }
|_{{\mbox{\scriptsize implicit} } } } \right|_{am_l = 0.0081}
\label{sverslvar}
\end{eqnarray}
where in the final step it is assumed that changes in the numerator and
denominator due to changing the quark mass in the Dirac operator from
$m_s $ to $m_l $ approximately cancel\footnote{Individually, one can
estimate the denominators in (\ref{substsl}) and (\ref{sverslvar}) to
differ by somewhat less than a factor of two, by comparing the right-hand
identity in (\ref{slderivs}) with the right-hand identity in (\ref{llderivs}),
corrected by the factor (\ref{implicvsfull}).}. This is the desired conversion
factor characterizing the strength of the implicit variation of $m_{res,l} $
with $m_s $ relative to the one with $m_l $. Correcting, as proposed above,
$\partial m_{res,l} / \partial m_l^{\mbox{\scriptsize bare} } $ by the
two factors (\ref{implicvsfull}) and (\ref{sverslvar}), one finally arrives
at the alternative estimate
\begin{equation}
\left.
\frac{\partial m_{res,l} }{\partial m_s^{\mbox{\scriptsize bare} } }
\right|_{am_s =0.081, am_l =0.0081} \approx 0.08 \cdot 0.14 \cdot (-0.020)
= -0.0002 \ ,
\end{equation}
consistent in order of magnitude with (\ref{simpest}), especially in view
of the latter being expected to represent an overestimate.

The estimate (\ref{simpest}) for
$\partial m_{res,l} /\partial m_s^{\mbox{\scriptsize bare} } $ at the
lowest light quark mass considered in the numerical calculations
in this work is used in section~\ref{scalrensec} to constrain the
influence of operator mixing effects in the renormalization
of $f_{T_s } $.

\end{document}